\documentclass[prd, twocolumn, superscriptaddress, nofootinbib, preprintnumbers]{revtex4-1}

\usepackage[bottom]{footmisc}
\usepackage{hyperref}
\usepackage[normalem]{ulem}
\usepackage[utf8]{inputenc}
\usepackage{amsmath}
\usepackage{mathrsfs}
\usepackage{mathtools}
\usepackage{graphicx}
\usepackage{braket}
\usepackage{color}
\usepackage[utf8]{inputenc}
\usepackage[T1]{fontenc}

\newcommand{\graphic}[2]{\includegraphics[width=#2\linewidth, type=pdf,ext=.pdf,read=.pdf]{#1}}
\newcommand{\lb}{\left (}
\newcommand{\rb}{\right )}
\newcommand{\blb}{\big (}
\newcommand{\brb}{\big )}
\newcommand{\Oc}{\Omega_{\mathrm{c}}}
\newcommand{\Ob}{\Omega_{\mathrm{b}}}
\newcommand{\Om}{\Omega_{\mathrm{m}}}
\newcommand{\OL}{\Omega_{\Lambda}}
\newcommand{\Onu}{\Omega_{\nu}}
\newcommand{\dl}{\delta_L}
\newcommand{\dw}{\delta_W}
\newcommand{\su}{\mathrm{su}}
\newcommand{\loc}{\mathrm{loc}}
\newcommand{\de}{\mathrm{DE}}
\newcommand{\dwsu}{\delta_{W, \su}}
\newcommand{\mpc}{\mathrm{Mpc}}
\newcommand{\msun}{\mathrm{M}_{\odot}}
\newcommand{\lcdm}{\Lambda\mathrm{CDM}}
\newcommand\lsim{\mathrel{\rlap{\lower4pt\hbox{\hskip1pt$\sim$}}\raise1pt\hbox{$<$}}}
\newcommand\gsim{\mathrel{\rlap{\lower4pt\hbox{\hskip1pt$\sim$}}\raise1pt\hbox{$>$}}}
\newcommand{\dee}{\mathrm{d}}
\newcommand{\brho}{\bar{\rho}}
\newcommand{\x}{\vec{x}}
\newcommand{\vk}{\vec{k}}
\newcommand{\pdf}{\mathscr{P}}
\newcommand{\pdfsu}{\pdf_{\su}}
\newcommand{\Rg}{R_{\mathrm{g}}}
\newcommand{\Rgs}{R_{\mathrm{g}\sigma}}
\newcommand{\Rgx}{R_{\mathrm{g}\xi}}
\newcommand{\rs}{r_\mathrm{s}}

\begin{document}
	
	\preprint{YITP-SB-2020-33}
	\title{The position-dependent matter density probability distribution function}
	\author{Drew Jamieson and Marilena Loverde \\{\it{\small  C.N. Yang Institute for Theoretical Physics, Department of Physics \& Astronomy, Stony Brook University, Stony Brook, New York, 11794, USA}}}
	
	\begin{abstract}
		We introduce the position-dependent probability distribution function (PDF) of the smoothed matter field as a cosmological observable. In comparison to the PDF itself, the spatial variation of the position-dependent PDF is simpler to model and has distinct dependence on cosmological parameters. We demonstrate that the position-dependent PDF is characterized by variations in the local mean density, and we compute the linear response of the PDF to the local density using separate universe N-body simulations. The linear response of the PDF to the local density field can be thought of as the linear {\em bias} of regions of the matter field selected based on density. We provide a model for the linear response, which accurately predicts our simulation measurements. We also validate our results and test the separate universe consistency relation for the local PDF using global universe simulations. We find excellent agreement between the two, and we demonstrate that the separate universe method gives a lower variance determination of the linear response.
	\end{abstract}
	
	\maketitle
	
	\section{Introduction}
		\label{sec:intro}
	
		One of the major scientific goals of cosmology is to observationally map the large-scale structure (LSS) of the Universe and use the statistics of density fluctuations in this map to constrain the parameters of physical models. Much progress will be made on the observational side of this program in the near future, with the development of several LSS surveys that will provide a wealth of data over the next decade \cite{Ivezic:2008fe,  Laureijs:2011gra, Spergel:2015sza, Aghamousa:2016zmz}. Interpreting and analyzing this data in ways that efficiently and rigorously constrain the parameter space of physical models remains a challenge. Partly, this is due to the nonlinear physics of gravitational clustering that drives cosmic structure formation.
		
		Nonlinear clustering leads to an important effect on the abundances of objects in the density field, for instance galaxies or voids, known as cosmic bias. Cosmic bias can be understood as the discrepancy between locally averaged quantities, such as the counts of galaxies in some finite region, and the corresponding globally averaged quantity. The bias arises because finite regions in the universe contain large-scale density perturbations, which cause the dynamics of structure formation to differ locally from the average clustering that occurs across the whole universe. Since the bias is due to dynamics and statistics of large-scale density modes, the bias also encodes information about the density field on large scales and provides a useful probe of those scales. 
		
		The concept of bias can be generalized to any observable that is affected by large-scale clustering. In this work, we study the local one-point statistics of the smoothed density field in finite subregions using N-body simulations. The presence of long-wavelength modes enhances clustering of overdense regions and reduces clustering of underdense regions, so information from the long-wavelength density perturbations is imprinted on the shapes of local one-point probability distributions. This effect can be interpreted as the bias of regions in the smoothed density field, selected based on their density.
		
		Pioneering observational work on the one-point statistics of galaxies was carried out by Hubble \cite{Hubble_1934}, and measurements of this observable have been made in more recent surveys \cite{Wild:2004me, Hurtado-Gil:2017dbm, Repp:2020kfd}. The one-point statistics of the matter field have also been measured using weak lensing maps \cite{Clerkin:2016kyr, Gruen:2017xjj}. Approaches to model the matter one-point statistics have been explored with N-body simulations \cite{Bouchet:1992hw, Kofman:1993mx, Gaztanaga:1999er, BetancortRijo:2001ge, Kayo:2001gu, Repp:2017znp, Shin:2017cwu, Klypin:2017jjg}. The first theoretical calculations of the matter field one-point statistics were based on perturbation theory \cite{Bernardeau:1994zd, Bernardeau:1994aq}, followed by nonperturbative approaches \cite{Valageas:2001zr, Ohta:2003zc, Lam:2007qw, Uhlemann:2015npz}. A sophisticated and accurate method for computing the matter field one-point statistics based on a path integral formalism was developed in \cite{Ivanov:2018lcg}, which we rely on extensively throughout this paper. Recent proposals have suggested using one-point statistics to detect primordial non-Gaussianity \cite{Friedrich:2019byw}, to constrain the sum of neutrino masses \cite{Uhlemann:2019gni}, and to obtain general cosmological parameter constraints \cite{Repp:2020etr}.
		
		We use separate universe simulations to study the response of the local, one-point probability distribution function (PDF) to the presence of large-scale density perturbations. In the separate universe approach, the long-wavelength modes are absorbed into the background cosmology, effectively altering the expansion history locally. Separate universe techniques have previously been used to study the local power spectrum and its response, which corresponds to the squeezed limit of the bispectrum \cite{Maldacena:2002vr, Creminelli:2004yq, Li:2014sga, Chiang:2014oga}. In the context of LSS, separate universe simulations have been used to measure the halo bias in simulations \cite{Li:2015jsz, Baldauf:2015vio}, which corresponds to the response of local halo mass functions. This method was used to study scale-dependent halo bias in cosmologies with massive neutrinos \cite{Chiang:2017vuk}, and in dynamical dark energy scenarios with both adiabatic \cite{Chiang:2016vxa} and isocurvature fluctuations \cite{Jamieson:2018biz}. Separate universe simulations have also been used to determine the bias of cosmic voids \cite{Chan:2019yzq, Jamieson:2019dmp}.
		
		The response of a local matter density PDF to the presence of a long-wavelength mode is not uniform as a function of density, indicating that some regions, or features of the cosmic web, are more sensitive to large-scale fluctuations than others. Since the statistics and dynamics of the large-scale modes are some of the most promising aspects of cosmology for constraining quantities such as primordial non-Gaussianity \cite{Meerburg:2019qqi}, the sum of neutrino masses \cite{Dvorkin:2019jgs}, and potentially the parameters of dynamical dark energy models \cite{Slosar:2019flp}, the responses of density PDFs could prove useful in targeting features of the density field for optimal parameter constraints. This idea is complementary to proposals for using marked correlation functions, in which the density field is nonlinearly transformed before correlation functions are computed  \cite{White:2016yhs,  Massara:2020pli}. It may be possible to use the responses of density PDFs to motivate optimal choices of nonlinear transformations for marked correlation functions.
	
		The outline of this paper is as follows. In Sec.~\ref{sec:PDFintro}, we define the Eulerian density PDF and the position-dependent, separate universe PDF. In Sec.~\ref{sec:model}, we review the details of a simple model for the PDF based on the evolution of isolated, spherical density perturbations. Using the separate universe formalism, we extend this model to a description of the local, position-dependent PDF and compute its linear response with respect to long-wavelength modes. We present the details of our simulations in Sec.~\ref{sec:sim} and describe our methods for estimating PDFs from the simulated density fields. In Sec.~\ref{sec:res}, we discuss the results of our separate universe simulations, including comparisons with model calculations. We present a validation of our separate universe responses by comparing with the cross-correlation between local, position-dependent PDFs and the matter density field measured in global universe simulations. We give examples of the position-dependent PDF's sensitivity to cosmological parameters in Sec.~\ref{sec:params}, and in Sec. \ref{sec:con}, we summarize our conclusions.

	\section{The PDF and the position-dependent PDF}
		\label{sec:PDFintro}

		% Definition and methods used on sim analysis
		\subsection{Global PDF}	
			\label{ssec:gpdf}
				
			We consider the PDF of the smoothed matter density field, $\rho_W$, where
			\begin{align}
			\label{eq:rhoWdef}
				\rho_W(t, \x) = \int \! \dee^3 y \ W(\vec{x} - \vec{y})\, \rho(t, \vec{y})\, ,
			\end{align}
			for a chosen window function $W$. In what follows, we will use a spherical top-hat window function of fixed radius $\rs$, given by
			\begin{align}
				W(\x - \vec{y}) = \frac{3}{4 \pi \rs^3} \Theta\big(\rs - |\x - \vec{y}\, | \big)\, .
			\end{align}
			Fluctuations of the spherically smoothed density field quantify the mass fluctuations within spheres of equal volume. The mean smoothed density is equal to the global mean density without smoothing, $\brho_W(t) = \brho(t)$, which follows from the normalization of the window function. 
			Defining 
			\begin{align}
				\label{eq:dw}
				1 + \dw = \frac{\rho_W}{\brho} \, ,
			\end{align}
			we denote the probability of finding a region where the density is between $1 + \dw$ and $1 + \dw + \dee\dw$ as
			\begin{align}
				\pdf(1 + \dw)\,\dee\dw\, . 
			\end{align}
			We will also refer to the above quantity as the {\em global }  PDF because it describes the probability of the density field reaching specific values in the Universe as a whole.
		
		\subsection{Position-dependent PDF}	
			\label{ssec:lpdf}
			
			In the same way that the abundances of objects such as halos, galaxies, or voids vary spatially, one also expects spatial variations in the PDF of the density field. Consider a local observer in a finite volume $V_{\su}$ located at position $\vec{x}$. This observer will see a local mean density, 
			\begin{align}
				\brho_{\su}(t, \x) = \frac{1}{V_{\su}} \int_{V_\su} \!\! \dee^3 y \ \rho(t, \vec{y}) \, ,
			\end{align}
			that differs from the global mean matter density $\brho$.
			We use the subscript $\su$ (as in ``separate universe'') to denote locally measured quantities in the region of $V_{\su}$. Observers within $V_{\su}$ will measure density contrasts with respect to the local mean density, 
			\begin{align}
				\label{eq:dloc}
				 \dwsu  =\ & \frac{\rho_{W, \su} -\brho_{\su}}{\brho_{\su}} \, ,
			\end{align}		
			where $\rho_{W, \su}$ is the density field smoothed on the same physical length scale as in Eq.~(\ref{eq:rhoWdef}). For a sufficiently large volume $V_{\su}$, the local mean density will differ from the global one by a small-amplitude density fluctuation $\dl$,
			\begin{align}
			\brho_{\su}(t, \x) = \brho(t) \blb 1+ \dl(t, \x) \brb \, .
			\end{align}
			We define the {\em position-dependent } PDF as the probability of finding a region of the density field in $V_{\su}$ with density between $1 + \dwsu$ and $1 + \dwsu + \dee \dwsu$. We denote this as
			\begin{align}
			\pdfsu(1 + \dwsu\, |\, \vec{x}) \, \dee \dwsu \, .
			\end{align}
			The function $\pdfsu$ trivially differs from $\pdf$ due to the difference in reference density in Eq.~(\ref{eq:dloc}), but, as we shall see, the background fluctuation $\delta_L$ also changes the evolution of structure in $V_\su$, leading to nontrivial differences between $\pdfsu$ and $\pdf$, which will be the main focus of this paper.
			
			While a local observer can only measure $\pdfsu$, an observer with access to a larger volume can extract correlations between $\pdfsu$ and fluctuations in the density field smoothed on scales $V_\su$, 
			\begin{align}
			\big\langle \pdfsu(1 + \dwsu\, |\, \x) \, \dl(\x') \big\rangle\,.
			\end{align}
			Our ansatz, which we will later verify, is that the spatial variation in $\pdfsu$ (that is, the $\x$ dependence) is entirely due to spatial fluctuations in the background density $\dl$. In this case, we can expand the local PDF as
			\begin{align}
			\label{eq:PDFexpand}
			\pdfsu \blb 1 + \dwsu \, | \, \dl(\x) \brb = \, &  \pdf(1 + \dw) + \frac{\dee \pdfsu }{\dee \dl} \dl(\x)  \nonumber \\
			& + \mathcal{O}(\dl^2) \,,
			\end{align}
			and write
			\begin{align}
			\label{eq:pdPDFcorr}
			\frac{ \big \langle \pdfsu(1 + \dwsu \, | \, \vec{x})\, \dl(\x') \big\rangle}{\big\langle \dl(\x) \, \dl(\x') \big\rangle} \simeq \frac{\dee \pdfsu }{\dee \dl}\,.
			\end{align}
			The fractional difference between the locally estimated PDF and the global PDF is characterized by a linear response to the long-wavelength perturbations,		
			 \begin{align}
			\label{eq:PDFresponse}
			\frac{\pdfsu(1 + \dwsu \, | \, \dl)}{\pdf(1 + \dw)} - 1 \simeq \frac{\dee \log \pdfsu }{\dee \dl} \dl 
			\end{align} 
		
			In what follows, we will study a model of the PDF from the literature. We will then use the model to compute the linear response of the PDF to the presence of long-wavelength modes. This allows us to predict $\pdfsu$, $\dee \log \pdfsu /\dee \dl$, and the observable correlation between the position-dependent PDF and the large-scale density fluctuations given in Eq.~(\ref{eq:pdPDFcorr}). We will verify our model calculations of $\dee \log \pdfsu /\dee \dl$ by comparing to separate universe simulations in Sec.~\ref{ssec:Rg}. In Sec.~\ref{ssec:Rc}, we will measure the position-dependent PDF directly and verify Eq.~(\ref{eq:pdPDFcorr}).  
		
		% Model calculation
		\section{Models for the PDF and the Position-Dependent PDF}
			\label{sec:model}	
			
			In the following subsections, we derive a model for the PDF and its linear response in the separate universe. We largely follow the reasoning laid out in \cite{Lam:2007qw}. We also rely heavily on the calculations presented in \cite{Ivanov:2018lcg}, in which the authors develop a more sophisticated and accurate model for the PDF than the one we use throughout this paper. We present our extensions of these models, using the separate universe formalism, which allows us to compute the linear response of the PDF to large-scale density perturbations. Our aim is to demonstrate that while predicting the full shape of the PDF requires a rather complicated calculation as in \cite{Ivanov:2018lcg}, the linear response can be accurately predicted by just the leading order contributions to the PDF, which are obtained from analytic, spherical collapse calculations. The only input required for our model calculations is the linear matter power spectrum, which we obtained from CLASS \cite{Blas:2011rf}, using the parameters listed in Table~\ref{tab:cos}.
	
			\begin{figure*}
				\graphic{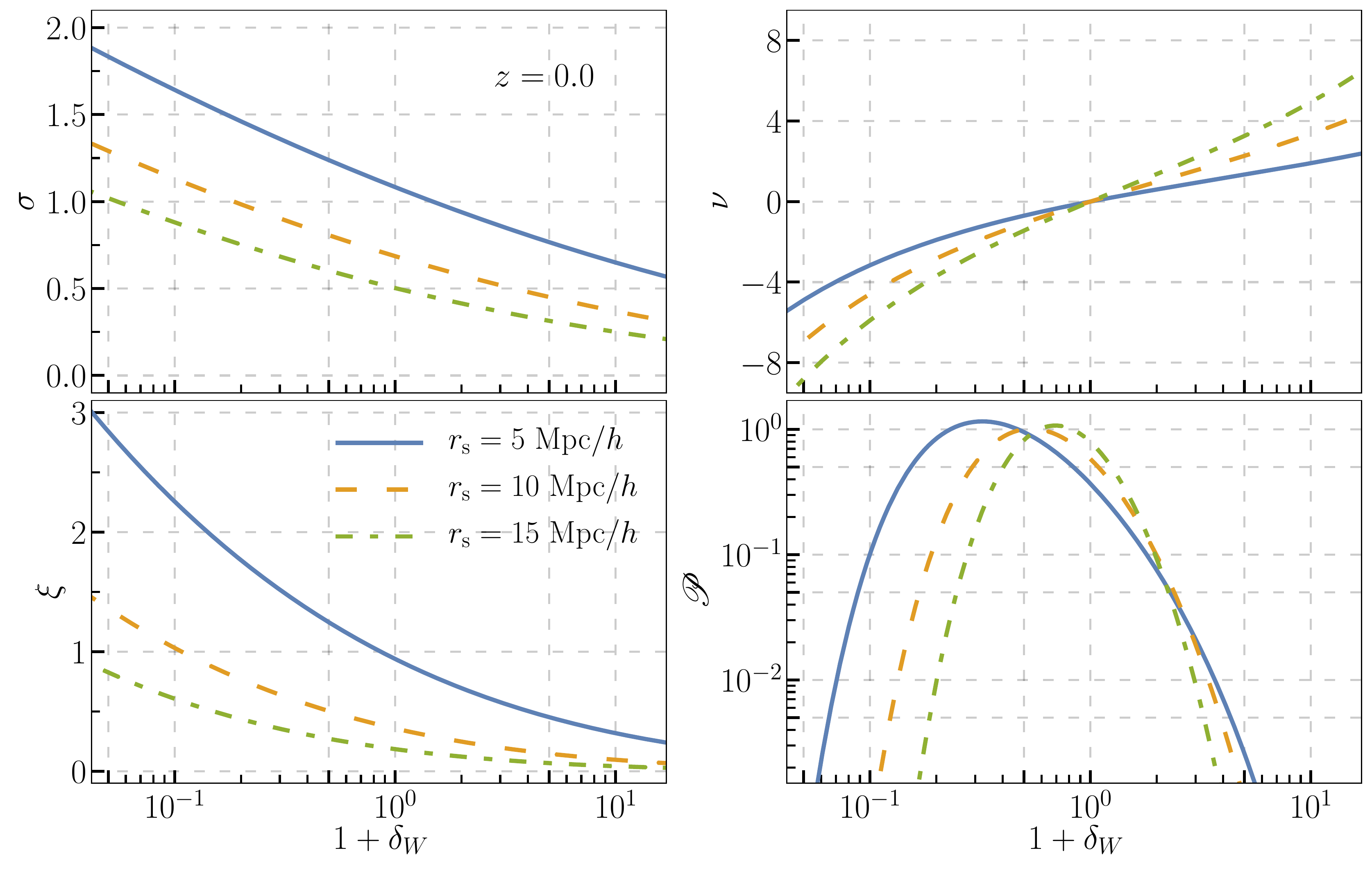}{1.}
				\caption{Quantities used to compute the Eulerian PDF model, evaluated at three smoothing scales, $\rs = 5~\mpc/h,\ 10~\mpc/h,\ \mathrm{and}\ 15~\mpc/h$, and redshift $z=0.0$. The top left shows the standard deviation of fluctuations in the smoothed matter field, defined in Eq.~(\ref{eq:sig2}) . The bottom left shows the smoothed spatial correlation function from Eq.~(\ref{eq:xi}). The top right shows the spherical collapse density map from Eqs. (\ref{eq:sc1})--(\ref{eq:nu}) . The bottom right shows the PDF model using Eqs.~(\ref{eq:P0})~and~(\ref{eq:A0}).}
				\label{fig:pdf_model}
			\end{figure*}
		
			\subsection{Spherical model for the global PDF}
				\label{ssec:PDFmodel}
		
				The PDF can be modeled by mapping the final density $1 + \dw(t_f)$ to an initial density $1 + \dw(t_i)$, where $t_i$ is chosen to be early enough so that the statistics of the density fluctuations are well approximated by a Gaussian distribution. Assuming that this map is deterministic and local \cite{Lam:2007qw}, we denote it
				\begin{align}
					\label{eq:Fdef}
					F \blb 1 + \dw(t_f) \brb \equiv \frac{D(t_f)}{D(t_i)} \dw(t_i)\, .
				\end{align}
				Here $D(t)$ is the linear growth factor, so the mapping provided by $F$ is between a final density and its corresponding initial density perturbation, which is linearly evolved to the final time. For isolated, spherically symmetric perturbations, this map is well approximated by the analytic solution in the Einstein--de Sitter (EdS) cosmology. For an initial overdensity, the parametric mapping is given by
				\begin{gather}
					\label{eq:sc1}
					1 + \dw(\theta) = \frac{9}{2} \frac{\blb \theta - \sin(\theta) \brb^2}{\blb 1 - \cos(\theta) \brb^3}\, ,\\
					\label{eq:sc2}
					F(\theta) = \frac{3}{20} \Big( 6 \blb \theta -\sin(\theta)  \brb \Big)^{2/3}\, ,
				\end{gather}
				and for an initial underdensity, we have
				\begin{gather}
					\label{eq:sc3}
					1 + \dw(\eta) = \frac{9}{2} \frac{\blb\sinh(\eta) - \eta \brb^2}{\blb \cosh(\eta) - 1 \brb^3}\, ,\\
					\label{eq:sc4}
					F(\eta) = -\frac{3}{20} \Big( 6\blb \sinh(\eta) - \eta \brb \Big)^{2/3}\, .
				\end{gather}
				The parameter in the underdense case takes values $-\infty \leq \eta < 0 $, and for the overdense case, $0\leq \theta<2\pi$. The map provided above is, to a good approximation, independent of both time and cosmology \cite{Bernardeau:1994zd}. At redshift $z$, we define the function $\nu$,
				\begin{align}
					\label{eq:nu}
					\nu (1 + \dw, z) \equiv \frac{F}{\sigma(R, z)}\, ,
				\end{align}
				where $\sigma(R, z)$ is the variance of the smoothed density field at the Lagrangian scale $R$, which is the comoving scale containing the mass
				\begin{align}
					M = \frac{4\pi}{3} \rs^3\, \brho\lb 1 + \dw \rb\, .
				\end{align}
				For a spherical density perturbation, the Lagrangian radius is given by,
				\begin{align}
					R = \rs\lb 1 + \dw \rb^{1/3}\, .
				\end{align}
				The variance of density fluctuations smoothed over spheres of radius $R$ is calculated as
				\begin{align}
					\label{eq:sig2}
					\sigma^2(R, z) = \int \frac{\dee^3k}{(2\pi)^3}\, \blb W(k R) \brb^2 P(k, z)\, , 
				\end{align}
				where $P(k, z)$ is the linear matter power spectrum and $W(k R)$ is the Fourier transform of the spherical top-hat window function,
				\begin{align}
					W(k R) = \frac{3}{(k R)^3} \blb \sin(k R) - k R \cos(k R) \brb\, .
				\end{align}
				While the spherical density map provided by $F$ is time independent, the linear power spectrum is proportional to the squared linear growth factor, $P \propto D^2$, which evolves with time. Thus, both $\sigma$ and $\nu$ are time and cosmology dependent, which leads to the time and cosmology dependence of the PDF model.
		
				To leading order in the spherical approximation, the PDF is given by the initial unit Gaussian distribution of $\nu$, multiplied by the Jacobian of the mapping provided by $\nu$ \cite{Bernardeau:2001qr},
				\begin{align}
					\label{eq:P0}
					\pdf_0(1 + \dw) = \frac{1}{\sqrt{2\pi}} \frac{\dee \nu}{\dee \dw} e^{-\nu^2/2}\, .
				\end{align}
				The Jacobian can be expressed as
				\begin{align}
					\frac{\dee \log \nu}{\dee \dw} = \frac{\dee \log F}{\dee \dw} +  \frac{1}{1+\dw}\lb 1 - \frac{\xi}{\sigma^2} \rb\, ,
				\end{align}
				where $\xi$ is the spatial correlation function at a distance $R$, smoothed over a spherical window of the same radius,
				\begin{align}
					\label{eq:xi}
					\xi(R, z) = \int \frac{\dee^3k}{(2\pi)^3}\,W(k R)\, \mathrm{sinc}(k R)\, P(k, z)\, .
				\end{align}
				The fact that the derivative of $\sigma$ is related to the smoothed spatial correlation function is a special property of the spherical top-hat window function. 
		
				While $\pdf_0$ is properly normalized by definition, it does not have the correct mean. A more general calculation of the PDF leads to the following form  \cite{Ivanov:2018lcg},	
				\begin{align}
					\label{eq:PDFAspher}
					\pdf = \mathscr{A} \pdf_0  \, ,
				\end{align}
				where the prefactor $\mathscr{A} $ is a function of $\dw$ and can be computed by considering fluctuations around the spherical collapse approximation. In order to recover the correct mean of the distribution, it is sufficient to consider only the leading order, spherically symmetric, or monopole fluctuations. At this order, the prefactor has the form
				\begin{align}
					\label{eq:A0}
					\log(\mathscr{A}_0) \simeq \dw\lb\frac{4}{21} - \frac{\xi}{\sigma^2}\rb + \mathcal{O}(\dw^2)  \, .
				\end{align}	
				In Fig.~\ref{fig:pdf_model}, we plot the quantities required to compute the spherical PDF model for a few different smoothing scales.
				
				Contributions from aspherical fluctuations can be computed by decomposing the fluctuations into multipole moments. However, the aspherical part of the prefactor is both independent of time and cosmology to a good approximation \cite{Ivanov:2018lcg} and thus does not contribute significantly to the spatial variations in the local PDF, which we discuss next.

			\subsection{Spherical model for the position-dependent PDF}
				\label{ssec:Rmodel}
				
				The growth of cosmic structure is locally modulated by the long-wavelength perturbations of the density field. In the separate universe formalism, this effect is described by defining a local cosmology in which the long-wavelength density modes, denoted $\dl$, are absorbed and treated as part of the homogeneous background density,
				\begin{align}
					\brho_\su= \brho \lb 1 + \dl \rb \, .
				\end{align}
				Then the local expansion history can be characterized by a separate universe scale factor and Hubble rate \cite{Sirko:2005uz}, which, to linear order, relate to the global expansion history,
				\begin{gather}
					\label{eq:t1}
					a_{\su} \simeq a \lb 1- \frac{1}{3} \dl \rb  \, , \\
					\label{eq:t2}
					H_{\su} \simeq H \lb 1 - \frac{1}{3} \dl' \rb  \, ,
				\end{gather}	
				where $\dl'$ indicates $\dee\dl/\dee\log a$.
	
				The effects of a long-wavelength mode can be determined by computing its evolution in the global cosmology with linear perturbation theory and then using the above equations to define its separate universe expansion history. The response of small-scale observables, from the linear to the deeply nonlinear regime, can be determined by running N-body simulations with a background expansion corresponding to the presence of a single long-wavelength density mode \cite{Li:2014sga, Li:2014jra, Baldauf:2015vio}. In particular, by running a pair of simulations corresponding to overdense and underdense regions, the linear response of an observable $\mathcal{O}_{\su}(a\, |\, \dl)$ in the separate universe can be determined by taking the finite difference derivative with respect to the long-wavelength mode,
				\begin{align}
					 \frac{\dee\log\mathcal{O}_{\su}}{\dee\dl}\bigg|_{\dl=0} \simeq \frac{\mathcal{O}_{\su}(a\,|+\dl) - \mathcal{O}_{\su}(a\,|-\dl) }{2\dl\mathcal{O}(a)}
				\end{align}
				where $\mathcal{O}(a)$ is the observable in the global universe. If the small-scale observable is taken to be the halo mass function in a region, then the separate universe response is the linear halo bias \cite{Li:2015jsz, Baldauf:2015vio}. Similarly, if the observable is taken to be the cosmic void size function measured in a region, then the separate universe response is the linear void bias \cite{Chan:2019yzq, Jamieson:2019dmp}. We will omit the notation indicating that the derivative is evaluated in the small $\dl$ limit.
	
				The model presented in the previous subsection describes the PDF as a functional over the linear matter power spectrum. The separate universe response of the PDF can therefore be computed in terms of the power spectrum response, given by \cite{Takada:2013wfa, Chiang:2014oga, Li:2014sga, Valageas:2013zda}
				\begin{align}
					\label{eq:RPk}
					\frac{\dee \log P_{\su}(k_{\su}\, |\, \dl)}{\dee \dl} = 2 \Rg - \frac{1}{3} \frac{\dee \log \blb k^{3} P(k) \brb}{\dee \log k}\, .
				\end{align}
				Here $P_{\su}(k_{\su} \, | \, \dl)$ is the local matter power spectrum measured by a separate universe observer in some finite region, and $k_{\su}$ is the comoving wave number defined with respect to the separate universe scale factor. Numerically equal comoving scales correspond to different physical scales in the separate and global universe coordinates because their scale factors take different values at equal times, according to Eq.~(\ref{eq:t1}).  We can interpret the second term on the right-hand side of Eq.~(\ref{eq:RPk}) as being due to the dilation of comoving scales between the separate universe and global cosmologies. Matching physical wave numbers, we have
				\begin{align}
					\label{eq:kphys}
					k_{\mathrm{phys}} = \frac{k}{a} = \frac{k_\su}{a_\su}\, ,
				\end{align}
				so that $k_{\su} \simeq k \lb 1 - \dl/3 \rb$. Similarly, to match spatial distances,
				\begin{align}
					\label{eq:rphys}
					r_{\mathrm{phys}} = a\,r = a_\su\, r_\su .
				\end{align}	
				The dilation contribution to the power spectrum response is nondynamical, in that it is trivially due to a coordinate change, and can be computed from the global power spectrum alone.
				
				The term $\Rg$ in Eq.~(\ref{eq:RPk}) is referred to as the growth response, and it represents the dynamical effect that a long-wavelength mode has on the local growth history. It is defined as the partial derivative of $\log (P_\su)$ with respect to $\dl$ at fixed comoving wave number,
				\begin{align}
					\Rg = \frac{1}{2} \frac{\partial \log P_\su}{\partial \dl}\bigg |_{k_\su = k}\, .
				\end{align}
				This derivative is taken between power spectra evaluated at different physical wave numbers, according to Eq.~(\ref{eq:kphys}).
	
				Suppose the mode $\dl$ has wave number $k_L$, and then the growth response also has an implicit dependence on $k_L$. In $\lcdm$, long-wavelength modes evolve according to the linear growth factor, which is independent of $k_L$. The growth response is therefore also independent of $k_L$, and we refer to this scenario as scale-independent growth. However, $\Rg$ can depend on $k$. 
				
				For $k$ in the quasilinear to nonlinear regime, the growth response $\Rg$ is $k$ dependent and must be computed from higher order perturbation theory or measured in simulations. In the low-$k$ limit, the term $\Rg$ is due to the local change in the linear growth factor,
				\begin{align}
					\lim_{k \rightarrow 0}\Rg = \frac{\dee \log D_{\su}}{\dee \dl} \, .
				\end{align}
				For EdS, this linear growth response is exactly $\Rg = 13/21$, while in $\lcdm$, $\Rg$ is well approximated by the EdS value; the onset of dark energy domination increases the growth response by less than a percent at redshift $z=0.0$. In more general cosmological scenarios, where $\dl$ has $k_L$-dependent evolution, $\Rg$ becomes scale dependent even in the linear regime, and it can differ significantly from the EdS value \cite{Chiang:2016vxa, Chiang:2017vuk, Jamieson:2018biz}.
	
				In the separate universe, we define the smoothed density fluctuations with respect to the local mean,
				\begin{align}
					1 + \dwsu = \frac{\rho_{W, \su}}{\brho \lb 1 + \dl \rb}\, ,
				\end{align}
				where $\rho_{W, \su}$ is the smoothed density in a spherical window function of radius $r_{\mathrm{s}, \su} = \rs \lb1 + \dl/3 \rb$, which is comoving with respect to the separate universe cosmology. 
	
				Calculating the local variance in matter density fluctuations from Eq.~(\ref{eq:sig2}), but using the local, separate universe power spectrum, the linear response of $\sigma$ becomes
				\begin{align}
					\frac{\dee \log \sigma_{\su}(R_{\su}\, |\, \dl)}{\dee \dl} = \Rgs + \frac{1}{3} \frac{\dee \log \sigma}{\dee \log \rs}\, .
				\end{align}
				The growth response of $\sigma$  is defined as
				\begin{align}
					\label{eq:Rgsig}
					\Rgs(R) = \frac{1}{\sigma^2(R)} \int \frac{\dee^3k}{(2\pi)^3}\, \blb W(k\, R) \brb^2 \, \Rg \, P(k)\, .
				\end{align}	
				The separate universe response of $\xi(R)$ is given by a similar expression,
				\begin{align}
					\frac{\dee \log \xi_{\su}(R_{\su}\, |\, \dl)}{\dee \dl} = \Rgx + \frac{1}{3} \frac{\dee \log \xi}{\dee \log \rs}\, .
				\end{align}
				In this case, the growth response of $\xi$ is
				\begin{align}
					\label{eq:Rgxi}
					\Rgx(R) = \frac{2}{\xi(R)} \int \frac{\dee^3k}{(2\pi)^3}\, W(k\, R)\, \mathrm{sinc}(k\, R) \, \Rg \, P(k)\, .
				\end{align}
				If $\Rg$ is $k$ independent then we have $\Rgx = 2 \Rgs = 2 \Rg$. Otherwise, $\Rgs$ and $\Rgx$ will depend on the smoothing radius $\rs$, which is expected when the smoothing is in the nonlinear regime ($\rs \lesssim 10~\mpc/h$ at redshift $z =  0.0$). This smoothing dependence will differ slightly between $\Rgs$ and $\Rgx$ because they are convolutions of the linear power spectrum with different functions.  If $\dl$ has scale-dependent evolution, then $\Rgs$ and $\Rgx$ will both have the same $k_L$ dependence, which they inherit directly from $R_{\mathrm{g}}$.

				\begin{figure}
					\graphic{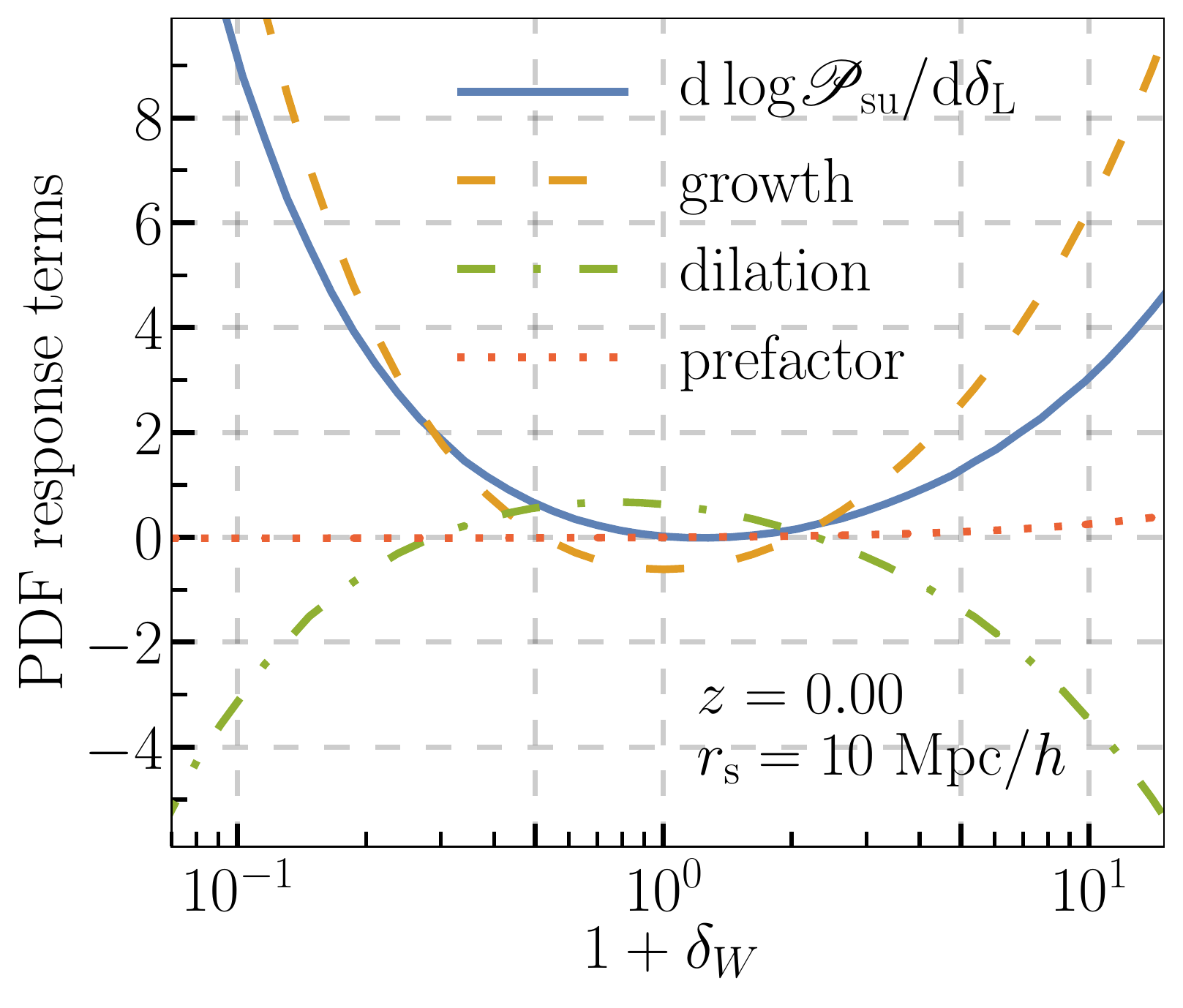}{1.}
					\caption{Decomposition of the PDF separate universe response, Eq.~(\ref{eq:RP}), from the spherical model calculation. The solid blue curve shows the full model prediction. The orange, dashed growth curve shows the contribution due to the change in the linear growth factor as in Eq.~(\ref{eq:Rgpdf}), which is a dynamical effect. The green, dash-dotted dilation curve is due to the difference between the separate universe and global comoving coordinates, given in Eq.~(\ref{eq:dPDFdrs}). The red dotted curve shows the dilation contribution from the spherical prefactor in Eq.~(\ref{eq:dAsphddL}).}
					\label{fig:Rpdf_terms}
				\end{figure}
	
				The linear response of the PDF is
				\begin{align}
				\label{eq:RP}
				\frac{\dee \log \pdf_{\su}(1 + \dwsu \, | \, \dl)}{\dee \dl}  
					= R_{\mathrm{g}\pdf} + \frac{1}{3} \frac{\dee \log \pdf}{\dee \log \rs} \, .
				\end{align}
				The first term represents the dynamical effect that a long-wavelength mode has on the one-point statistics of the smoothed density field, and we refer to it as the growth response of the PDF. This term can be thought of as the partial derivative of  $\log\lb\pdfsu\rb$ with respect to the long-wavelength mode at fixed comoving scale,
				\begin{align}
					\label{eq:pdR}
					R_{\mathrm{g}\pdf} 	= \frac{\partial \log \pdf_{\su}(1 + \dwsu \, | \, \dl)}{\partial \dl}\bigg|_{r_{\mathrm{s},\su} = \rs}  \, .
				\end{align}	
				This derivative is taken between PDFs of the density field smoothed over different physical scales, according to Eq.~(\ref{eq:rphys}).
	
				In the spherical model, the leading-order PDF growth response is
				\begin{align}
					\label{eq:Rgpdf}
					R_{\mathrm{g}\pdf}  = &  \lb \nu^2 - 1 \rb \Rgs +  \frac{\xi}{\sigma^2} \frac{\dee \log\dw }{\dee \log \nu} \lb 2 \Rgs - \Rgx \rb  \, ,	
				\end{align}
				where we have used the fact that $F$ is, to a good approximation, cosmology independent.
				If the growth responses $\Rgs$ and $\Rgx$ do not depend on $\rs$, the second term vanishes and we have $R_{\mathrm{g}\pdf}  = \lb \nu^2 - 1 \rb R_{\mathrm{g}}$, which is expected for $\lcdm$ when the smoothing scale is not too deep into the nonlinear regime. We will see that this is quite accurate in $\lcdm$ even when the smoothing radius approaches the nonlinear scale. 
	
				At leading order, the dilation term of the PDF response is
				\begin{align}
					\label{eq:dPDFdrs}
					\frac{\dee \log \pdf_0}{\dee \log \rs} &  =  \lb \nu^2 - 1 \rb \frac{\dee \log \sigma}{\dee \log \rs}  \nonumber \\
					& \ +  \frac{\xi}{\sigma^2}\frac{\dee \log \dw }{\dee \log \nu} \lb 2  \frac{\dee \log \sigma}{\dee \log \rs} -  \frac{\dee \log \xi}{\dee \log \rs} \rb  \, .
				\end{align}
				The leading correction from the prefactor in Eq.~(\ref{eq:PDFAspher}) contributes additional terms to the PDF response,
				\begin{align}
					\label{eq:dAsphddL}
					\frac{\dee \log \mathscr{A}_{0, \su}}{\dee \dl} =  \dw \frac{\xi}{\sigma^2} \lb 	2\frac{\dee \log  \sigma_{\su}}{\dee \dl}  -  \frac{\dee \log \xi_{\su}}{\dee \dl} \rb \, .
				\end{align}
				When the growth response is independent of $\rs$, the spherical prefactor only contributes to the dilation part of the PDF response. As shown in Fig.~\ref{fig:Rpdf_terms}, this contribution is negligibly small everywhere except in the high density tail of the PDF where $\dw \gtrsim 10$. While the leading correction recovers the shape of the PDF near its peak, at this order the model fails to reproduce the distribution's tails, so the dilation computed from this model is not expected to be accurate. However, as will be shown below, this model actually does describe the dilation at high densities well. On the other hand, the dilation can easily be obtained directly from data, by measuring the global PDF at different smoothing scales.
	
				The model prediction for the PDF response is plotted in Fig.~\ref{fig:Rpdf_terms}, along with the individual contributions from the leading growth, dilation, and subleading dilation due to the prefactor. An overdense region will have enhanced clustering, which increases small-scale overdensities and further depletes matter out of the small-scale underdensities. The net effect is to increase the occupation of the distribution's tails, so the PDF is wider. Similarly, an underdense region has reduced clustering, meaning more regions have densities near the peak of the PDF and the tails are diminished. This can clearly be seen in Fig.~\ref{fig:Rpdf_terms}, where the PDF growth response is positive throughout the low and high density tails of the distribution, and there is a slightly negative response near the peak.
				
				Although the model does not accurately describe the effect of scale dilation except at high densities, the qualitative results are informative. In order to match the physical smoothing scale in the global universe, the overdense separate universe has a slightly larger comoving smoothing scale, while the underdense separate universe has a slightly smaller comoving smoothing scale. Smoothing on larger comoving scales yields smaller amplitude density fluctuations, which in turn leads to a narrower PDF, while smoothing on smaller comoving scales broadens the distribution. This effect can be seen in Fig.~\ref{fig:Rpdf_terms} from the dilation term, which is negative in the tails and positive near the peak. The prefactor term is also shown and gives a small positive contribution to the response only in the extremely high density tail. 

	\section{Simulations}
		\label{sec:sim}
			
		We ran three sets of 20 N-body simulations. One set corresponded to the global expansion history under a $\lcdm$ cosmology. The other two simulation sets had expansion histories corresponding to separate universes in overdense and underdense regions. The long-wavelength mode was normalized to $\dl = 0.01$ at redshift $z=0.0$. For all three sets, we used the same set of 20 random seeds to generate the initial conditions, which ensured cosmic variance cancellation occurs at leading order when computing separate universe response observables. Our simulations were run using a modified version of Gadget2 \cite{Springel:2005mi}, which reads in and interpolates from tabulated values of the separate universe scale factor and Hubble rate, rather than integrating the Friedmann equation to compute the expansion history. Our simulation box size was chosen to be $L_{\mathrm{box}} = 1000\ \mathrm{Mpc}/h$ with $N_\mathrm{p} = (1024)^3$ dark matter particles. The particle mass was $M_\mathrm{p} = 1.108\times 10^{11}\ \msun$. These simulations were previously used for a study of void bias in the separate universe \cite{Jamieson:2019dmp}. The simulation parameters are summarized in Table~\ref{tab:cos}.
				
		Note that the comoving sizes of the simulation boxes are all equal, and so is the particle number per simulation, so the particle masses are the same in the separate universe and global universe simulations. While the mean comoving density in every simulation box is the same, the physical volume of the overdense simulation is actually smaller than the corresponding global simulation by a factor of $(1-\dl)$, so it is in fact an overdense box. By running simulations at fixed comoving size, we isolated the dynamical, growth response from separate universe response observables \cite{Li:2014sga, Wagner:2014aka}. This required modifying the output times of our separate universe simulations according to Eq.~(\ref{eq:t1}), in order to match with the time in the global coordinates. The dilation term in separate universe responses can be measured in global universe simulations, as will be described below.
		
		Snapshots from our simulations containing the dark matter particle positions and velocities were saved at redshifts $z=0.0$, $z=0.5$, and $z=1.0$. The number density of dark matter particles was then sampled on a grid of $N_\mathrm{s} = 512^3$ overlapping spheres. These densities were converted to values of $1+\dwsu$ in our separate universe simulations, and $1+\dw$ in the global universe simulations simply by dividing by the mean number density of particles, $\bar{n} = \bar{n}_{\su} = 1.07~(\mpc/h)^{-3}$. We chose to consider the smoothing radius $\rs=10~\mpc/h$, which is approximately the nonlinear scale at redshift $z=0.0$. 
		
		\begin{table}
			\begin{tabular}{c c}
				\hline\hline
				Parameter & Value \\
				\hline
				$\OL$ & 0.7 \\			
				$\Om$ & 0.3 \\
				$\Ob$ & 0.05 \\
				$h$ & 0.7 \\
				$n_\mathrm{s}$ & 0.968 \\
				$A_\mathrm{s}$ & 2.137$\times10^{-9}$ \\ 
				$N_\mathrm{p}$ & $(1024)^3$ \\
				$L_{\mathrm{box}}$ & $1\ \rm{Gpc}/$$h$ \\
				$M_\mathrm{p}$ & $1.108\times10^{11}\ \msun$\\
				\hline
				\hline
			\end{tabular}
			\caption{Cosmological and N-body simulation parameters.}
			\label{tab:cos}
		\end{table}
		
		From our samples of the spherically smoothed density field, we computed the PDF using a kernel density estimator (KDE). This technique is widely used in statistics and data science, although it is less commonly used in cosmology. KDEs are superior to binning because the trade-off between variance and statistical bias is less severe, and KDEs have better convergence properties than histograms.
		
		Our approach was as follows. First, we converted the list of densities to a list $\big\{\!\log(1+\dw{}_{i})\big\}$ and estimated the PDF as
		\begin{align}
			\pdf_{\log} \blb( \log( 1 + \dw ) \brb = \frac{1}{w N_{\mathrm{s}}} \sum_{i = 1}^{N_\mathrm{s}} K \Biggl(\frac{1}{w} \log \lb \frac{1 + \dw}{1+\dw{}_{i}} \rb \Biggr) \, ,
		\end{align}
		where the kernel $K$ is a Gaussian function with unit variance and vanishing mean. The parameter $w$ is the kernel width and controls the level of smoothing over the discrete data samples. The estimated PDF was then converted back from $\log$-densities,
		\begin{align}
			\pdf(1 + \dw) = \frac{\pdf_{\log} \blb \log ( 1 + \dw ) \brb}{ 1 + \dw} \, .
		\end{align}
		The  density PDFs have long tails at high densities and fall off sharply at low densities, so the shape of the PDF is at least qualitatively reminiscent of a $\log$-normal distribution \cite{Coles:1991if, Bernardeau:1994aq}, which motivates our procedure.
		
		The choice of kernel width is important since a width that is too narrow will produce excess variance in the poorly sampled tails of the distribution, while a width that is too wide introduces statistical bias in regions where the curvature of the distribution is large. Assuming that the distribution being estimated by a KDE is itself a Gaussian, and taking the asymptotic limit of large sample size, the optimal width that minimizes the integrated squared error is \cite{Silverman_1986}
		\begin{align}
			w_{\mathrm{G}} = \sigma_{\pdf} \lb \frac{4}{3 N_\mathrm{s}} \rb^{1/5}\, .
		\end{align}
		Here, $\sigma_{\pdf}$ is the standard deviation of the true distribution, which can be estimated from the sample. Our matter density PDFs have longer tails at high densities than a $\log$-normal distribution, so the optimal kernel width is wider than $w_{\mathrm{G}}$. We measured the PDFs from our global simulations with $w = 1.5\, w_{\mathrm{G}}$, which we settled on by using leave-one-out cross-validation over the 20 PDFs measured in our set of simulations.
		
		The optimal width for estimating derivatives of a distribution, which is needed to compute $R_{\mathrm{g}\pdf}$, is not the same as the optimal width for the distribution itself. The optimal with for the first derivative scales with sample size as $N_\mathrm{s}^{-1/7}$ rather than $N_\mathrm{s}^{-1/5}$, so estimating derivatives of a PDF requires a wider kernel. This is also true for the separate universe responses, which are computed as finite difference derivatives. For estimating derivatives of the PDFs we used a kernel width of $w = 2.5\, w_{\mathrm{G}}$, which was also chosen by cross-validation. 
		
		\section{Results}
			\label{sec:res}
		
			\begin{figure*}
				\graphic{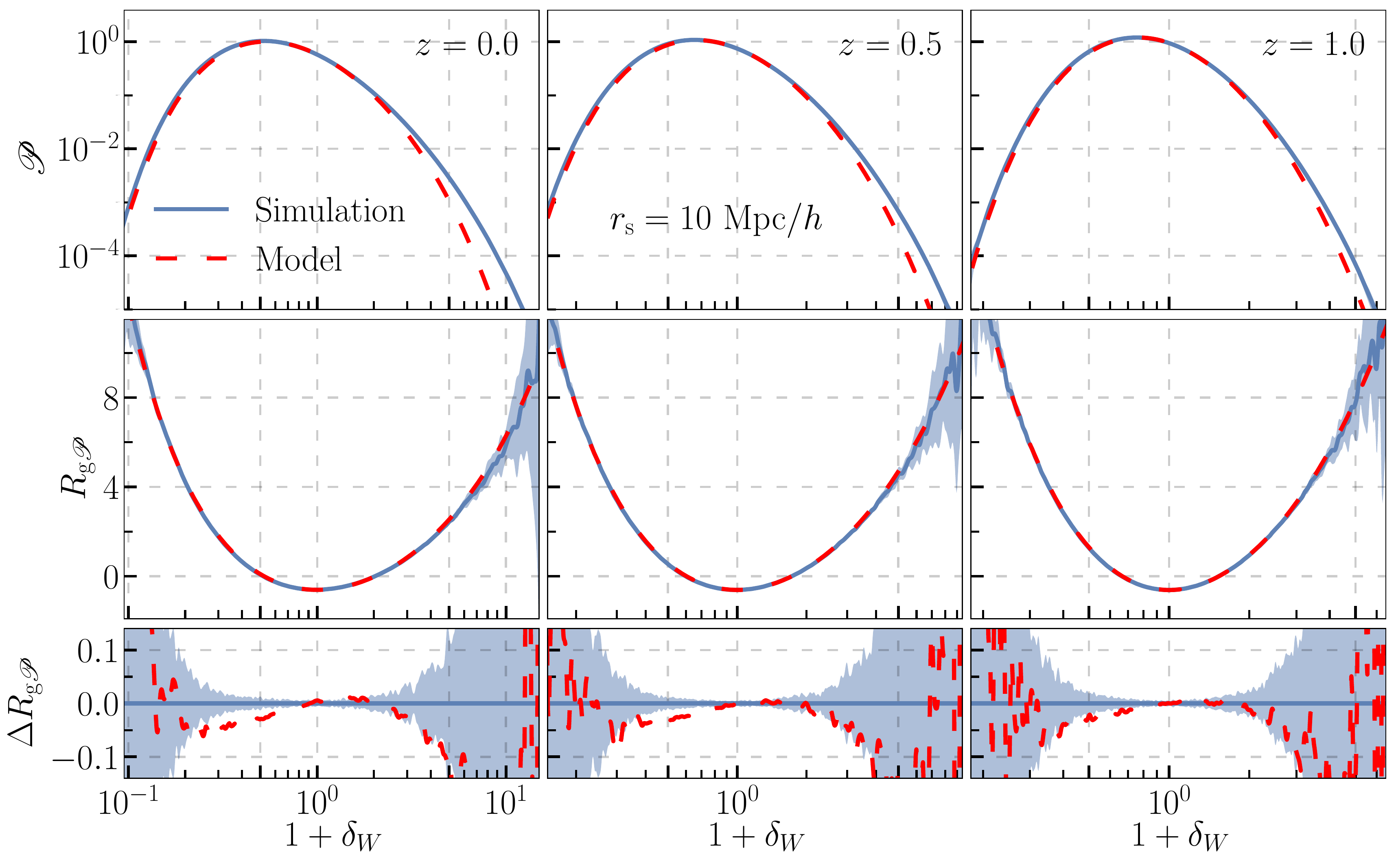}{1.}
				\caption{Top: The global PDF  of the density field smoothed with spherical top hats of radius $\rs = 10~\mpc/h$. Solid blue curves show the simulation results, with the shaded area indicating the $1\sigma$ bootstrap error, while the dashed red curves show the model prediction described in Sec.~\ref{ssec:PDFmodel}. Middle: The dependence of the PDF on the local mean density, as measured via the separate universe response of the PDF, defined in Eq.~(\ref{eq:pdR}). The dashed red curves show the model prediction for the same quantity, given in Eq.~(\ref{eq:Rgpdf}). Bottom: Difference between the separate universe response of the density PDF from simulations and the model prediction.}
				\label{fig:epdfs}
			\end{figure*}
		
			The PDFs measured in our simulations are shown in Fig.~\ref{fig:epdfs}, with shaded regions corresponding to $1\sigma$ bootstrap errors. The model calculations, which are also shown in Fig.~\ref{fig:epdfs}, were computed from $\pdf = \mathscr{A}_0 \pdf_0$, where $\pdf_0$ is given by Eq.~(\ref{eq:P0}), and the prefactor $\mathscr{A}_0$ is given by Eq.~(\ref{eq:A0}). As expected, the model accurately recovers the peak of the distribution but falls off too quickly in the tails, which is consistent with previous work \cite{Ivanov:2018lcg}. The model significantly improves with increasing redshift, as the measured distribution gets narrower. This is consistent with the time independence of the aspherical corrections to the PDF model. While these correction factors are constant with respect to time, the spherical model produces a much narrower distribution at higher redshifts, so the PDF does not have very strong support where the aspherical part of the prefactor becomes important \cite{Ivanov:2018lcg}.
	
		\subsection{The growth response of the PDF}
			\label{ssec:Rg}
		
			The growth response, measured from our separate universe simulations, is also shown in Fig.~\ref{fig:epdfs} and compared with the model calculation from Eq.~(\ref{eq:Rgpdf}). The model does an excellent job predicting the shape of the growth response, even at extreme densities where the model poorly predicts the shape of the PDF itself. The accuracy of the model is consistent between redshifts $z=0.0$ and $z=1.0$ for mid to high densities with $\dw>0.7$. At low densities, $\dw < 0.7$, the model is slightly worse at redshift $z = 0.0$ than at earlier redshifts. We have included plots of the difference between the model prediction and the measured response in Fig.~\ref{fig:epdfs}. We have plotted the difference, rather than the fractional difference, because of the zero crossings of the response. 
		
			At redshift $z=0.0$, the differences between our predicted and measured responses is $\Delta R_{\mathrm{g}\pdf}\simeq-0.05$ at $1+\dw\simeq0.2$, which is a discrepancy of 0.6\%. At $1+\dw\simeq5$ we have $\Delta R_{\mathrm{g}\pdf}\simeq-0.1$, which is a 3\% discrepancy. The model predictions at this redshift are $1\sigma$ from the simulation measurements for $\dw >3$, while the model is $2\sigma$ from the simulation result between $\dw=0.3$ and $\dw=0.6$. The model comparisons are similar for redshift $z=0.5$, although the range of low densities at which the model is discrepant by $2\sigma$ is narrower, spanning from $\dw=0.4$ to $\dw=0.6$. The model is accurate at $1\sigma$ for both high and low densities at redshift $z=1.0$.
	
			\begin{figure*}
				\graphic{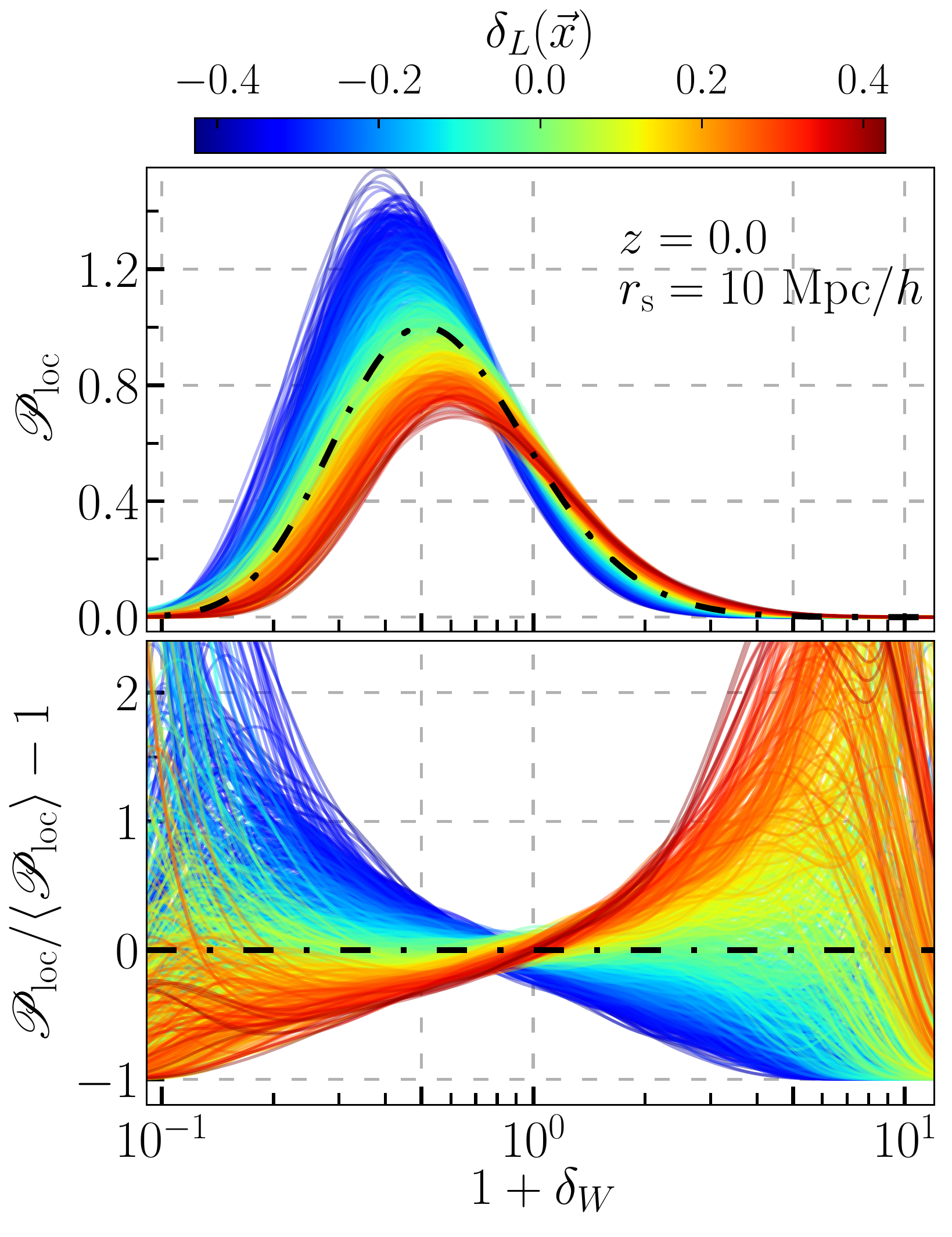}{0.482}
				\graphic{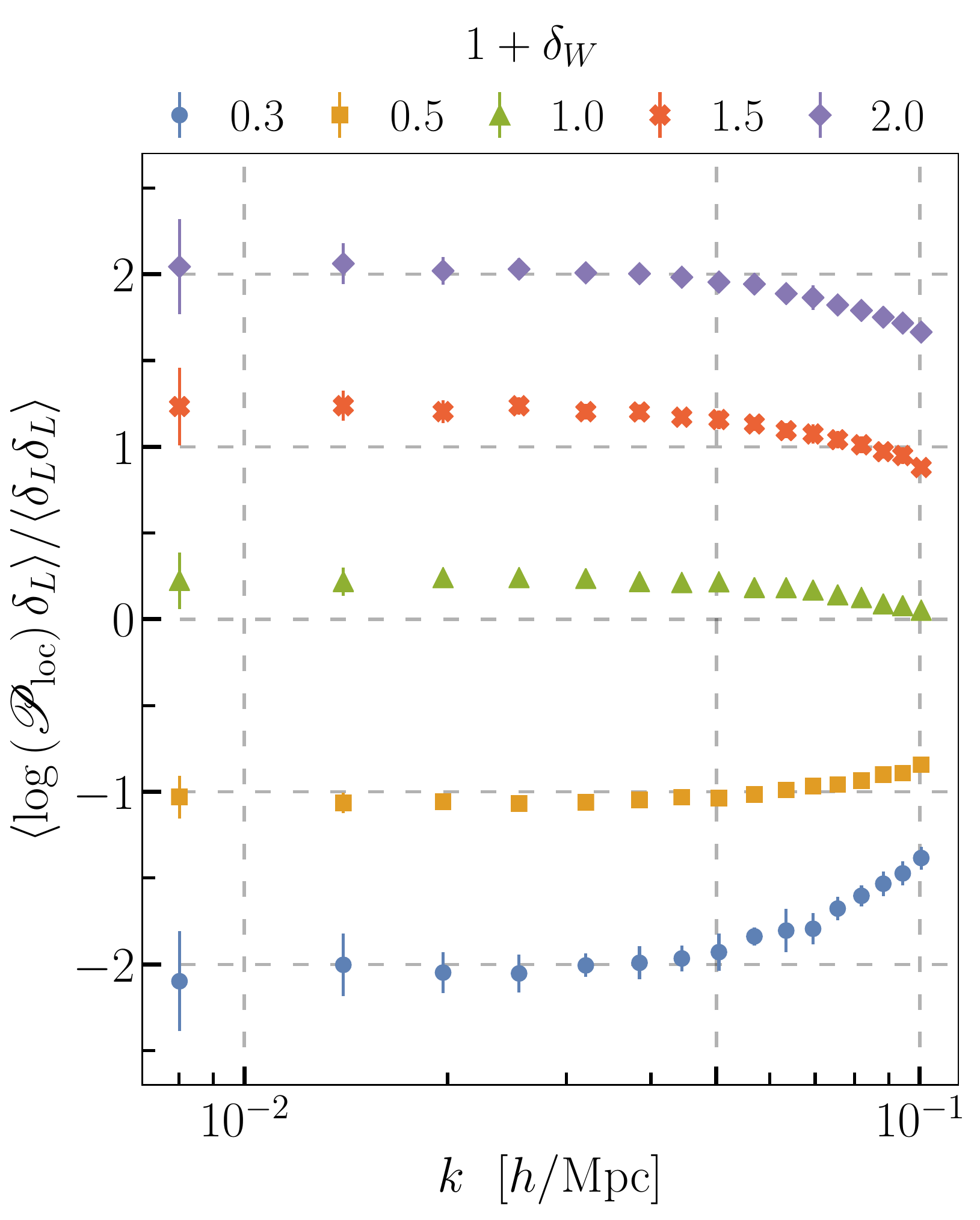}{0.495}
				\caption{Left: The local PDF measured in subboxes of global universe simulations. Small-scale density contrasts are defined with respect to the global mean as in Eq.~(\ref{eq:dw}), and the color indicates the local, mean density in the subbox, $\brho\lb 1 + \dl \rb$. The black, dash-dotted curve shows the global PDF. The bottom panel shows the fractional difference between the local PDFs and the global mean PDF. Right: The cross power spectrum between the fluctuations of the local PDFs and the large-scale density fluctuations in each subbox, divided by the matter power spectrum of the large-scale density fluctuations. This is effectively the {\em bias factor} of the PDF at fixed values of $\dw$.}
				\label{fig:local_clustering}
			\end{figure*}
	
			These results are also consistent with the aspherical corrections of the PDF model being independent of cosmology. Although the spherical model fails to reproduce the tails of the distribution, this is due to the omission of a prefactor that depends on $\dw$ but has no strong dependence on cosmological parameters. On the other hand, since the prefactor is the same from the point of view of observers in overdense and underdense separate universes, the aspherical prefactor does not contribute to the PDF response and in fact cancels when taking the $\log$-derivative in Eq.~(\ref{eq:pdR}). Notice, however, that the aspherical prefactor is sensitive to the smoothing scale, and so it does contribute to the dilation response, which we discuss shortly.
			
		% Clustering method and validation
		
		\subsection{Validation in the global universe}
			\label{ssec:Rc}
		
			\begin{figure*}
				\graphic{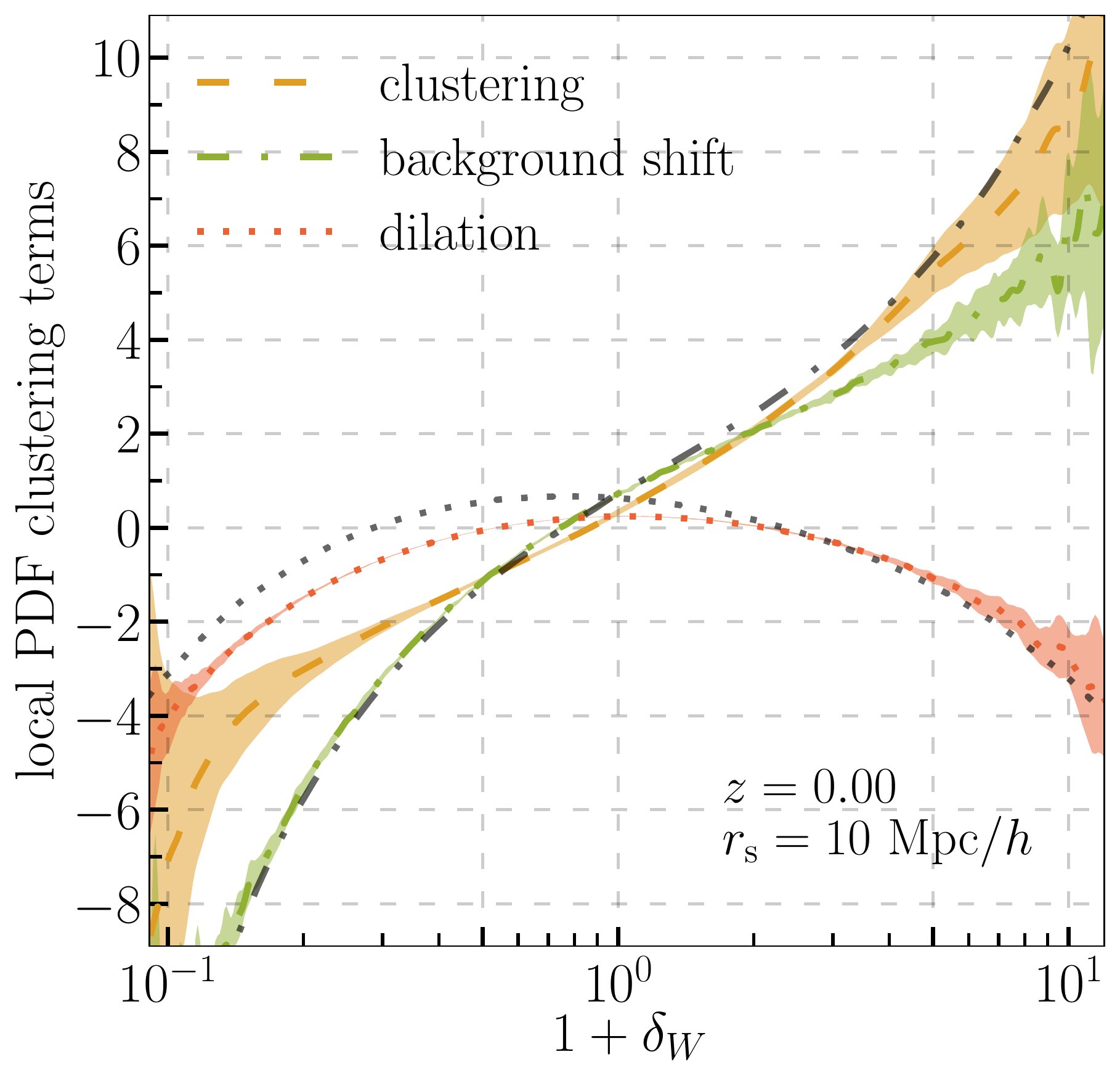}{0.482}
				\graphic{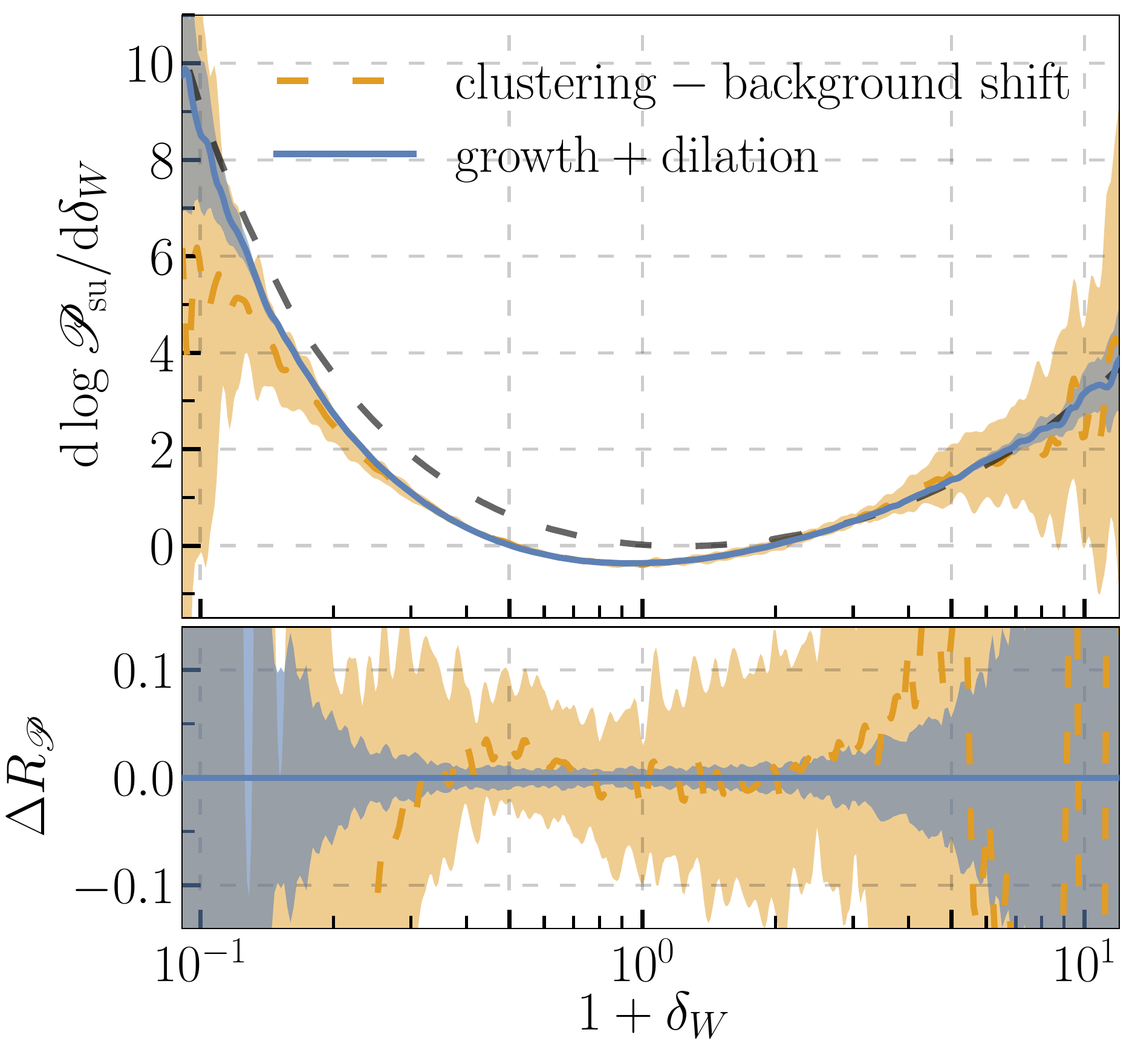}{0.498}
				\caption{Left: The local PDF clustering bias, as given by the low-$k$ limit of the left-hand side of  Eq.~(\ref{eq:Rloc}). Also plotted are the nondynamcial contributions to the PDF clustering bias decomposition given in Eq.~(\ref{eq:Rlocdecomp}) from the shift in the mean density and the dilation of comoving scale, both measured in simulations with the global cosmology. The gray curves show the model predictions for the individual terms in the decomposition of the clustering bias. Right: Comparison between the response of the PDF measured in separate universe simulations and adding the nondynamical dilation term, and the PDF response measured in simulations with the global cosmology using the clustering method and subtracting the background shift contribution. The agreement between the two validates our separate universe measurements, and the ansatz that the spatial variation in the PDF is generated by variations in the local mean density. The bottom panel shows the PDF response minus the separate universe measurement of the PDF response. All shaded regions correspond to $1\sigma$ bootstrap errors.}
				\label{fig:cluster}
			\end{figure*}
		
			Using our global universe simulations, we measured the PDF in large subboxes in order to validate our separate universe results and test the consistency relation from Eq.~(\ref{eq:RP}). It is possible to do this directly, by computing the position-dependent separate universe PDFs within large subboxes. However, we find that the responses obtained in this way have a large variance. Instead, we consider the local PDFs that a global observer would measure within each subbox, which greatly reduces the variance of the measured responses.
			
			The local  PDF , which is measured with respect to the global mean density, is related to the separate universe PDF measured within the same region,
			\begin{align}
				\frac{\pdf_{\loc}(1 + \dw \, | \, \dl)}{\pdfsu(1 + \dw \, | \, \dl)}
				\simeq 1 - & \dl \lb 1 + \frac{\dee \log \pdf}{\dee \log (1 + \dw)} \rb \,  .
			\end{align}	
			In this expression, both $\pdf_{\loc}$ and $\pdfsu$ are evaluated at the same density, $1 + \dw$, as is the derivative of the global PDF on the right-hand side. 
		
			We define the local PDF as a function of position, $\pdf_{\loc}(1 + \dw | \, \x)$, which is measured in a cubic subbox centered on comoving position $\x$. The local PDF fluctuations can be spatially Fourier transformed into modes denoted $\log\pdf_{\loc}(1 + \dw | \, \vk)$. For brevity, we will denote the local fluctuations in the PDF $\pdf_{\loc} / \pdf - 1 \simeq \log(\pdf_{\loc} / \pdf)$, although in practice we actually compute the fractional difference. Cross-correlating, or taking the cross power spectrum between the local PDF fluctuations and the large-scale density modes of the subboxes, gives
			\begin{align}
				\label{eq:Rloc}
				\frac{\big \langle \log \pdf_{\loc}(\vk )\, \dl(\vk') \big\rangle}{\big \langle \dl(\vk) \,  \dl(\vk') \big\rangle} = - & \lb 1 + \frac{\dee \log \pdf}{\dee \log (1 + \dw)} \rb  \nonumber \\ & + \frac{\dee \log \pdfsu}{\dee \dl}\, .
			\end{align}
			We refer to this as the local PDF clustering. The first term is due to the shift in the background density of the subboxes compared to the global universe, while the second term is the separate universe response, which according to Eq.~(\ref{eq:RP}) contains both the growth and dilation terms. The background shift term, like the dilation, is nondynamical and can be determined from the global PDF alone. To summarize, we anticipate that the local PDF clustering can be decomposed as follows,
			\begin{align}
				\label{eq:Rlocdecomp}
				\frac{\big \langle \log \pdf_{\loc}(\vk )\, \dl(\vk') \big\rangle}{\big \langle \dl(\vk) \,  \dl(\vk') \big\rangle} = - &  \underbrace{\lb 1 + \frac{\dee \log \pdf}{\dee \log (1 + \dw)} \rb}_{\textrm{ background\,\, shift}}  \\ 
				& + \underbrace{\frac{\partial \log \pdf_{\su}}{\partial \dl}}_{\textrm{growth}} \bigg|_{r_{\mathrm{s},\su} = \rs}\nonumber\\
				& +		 \underbrace{\frac{1}{3} \frac{\dee \log \pdf}{\dee \log \rs}}_{\textrm{dilation}}\,.		\nonumber
			\end{align}
			
			We measured the local PDF on a grid of $512^3$ overlapping cubic subboxes with side length of $250~\mpc/h$. Within each subbox,  we distributed the particles among a mesh of $256^3$ overlapping spherical top hats of radius $\rs = 10~\mpc/h$ and computed the density fluctuations, as defined in Eq.~(\ref{eq:dw}). We then compute the kernel density estimator of the local PDF for each subbox. In total, including the 20 independent realizations of the initial conditions that we simulated, we obtained $7.86\times10^5$ samples of the local density PDF. 
		
			A sample of 500 local PDFs, uniformly distributed according to the local subbox density $\dl$, is shown in Fig.~\ref{fig:local_clustering}. The PDFs in underdense subboxes (blue curves) are clearly narrower, or more sharply peaked, while the overdense subboxes (red curves) have broader distributions. Notice that, unlike the separate universe PDF response, the fractional difference between the local and global PDF shows a strong, negative response in the underdense tail. That is, underdense spheres are much more abundant in underdense subboxes. This is due to the background shift, which is not present in the separate universe PDF. In fact, this strong, negative response corresponds exactly to the peak of the distribution being shifted to lower densities for the blue curves, while it moves to higher densities for the red curves. For the separate universe PDFs, the position of the peak is affected considerably less by the presence of $\dl$, and the positive response in the negative tail in that case is due to gravitational clustering.
				
			The Fourier transformations of the local density PDF and the large-scale density fluctuations in the subboxes were computed using FFTW3 \cite{FFTW05}. The local PDF clustering at redshift $z=0.0$ is plotted in Fig.~\ref{fig:local_clustering} as a function of wave number for a range of small-scale densities. The PDF clustering is clearly scale independent on the largest scales. We fit the $k$ dependence with a linear polynomial in $k^2$ below $k_{\mathrm{max}}=0.05~h/\mpc$. The constant term from the fit gives the linear, long-wavelength limit of the PDF clustering, which should differ from our separate universe responses by exactly the background shift term in Eq.~(\ref{eq:Rloc}). In order to test this, we measured the background shift from the global PDF's derivative with respect to $1 + \dw$, $\dee \log \pdf / \dee \log (1 + \dw)$. We also measured the dilation term, $\dee \log \pdf / \dee \log \rs$, by estimating the global PDF at five smoothing scales ranging from 1\% smaller to 1\% larger than $\rs = 10~\mpc/h$, and fit the $\rs$ dependence with a quadratic polynomial.
		
			The linear limit of the clustering measured in our global universe simulations at redshift $z=0.0$ is plotted in Fig.~\ref{fig:cluster}, along with our measurements of the background shift and the dilation terms. The nondynamical terms in the response were also computed using the PDF model, and these are also shown in Fig.~\ref{fig:cluster}. The model prediction for the dilation term is inaccurate below $\dw = 3$, while at higher densities it agrees with our simulations. The model prediction for the background shift term is inaccurate for $\dw > 1.5$, while it reproduces our simulation results well at low densities. Unfortunately, this means that the simple, spherical model prediction for the local PDF clustering is inaccurate at all densities. However, the background shift and dilation are measured with small variance, so we can use the simulation data to directly test the separate universe consistency relation for the PDF response given in Eq.~(\ref{eq:PDFresponse}).
					
			In Fig.~\ref{fig:cluster}, we show a plot comparing two methods for obtaining the separate universe response of the PDF.  For the first method, we subtracted the background shift from the local clustering. For the second method, we added the dilation term to the growth response measured in separate universe simulations. The two methods are in agreement across the full range of densities. The separate universe method achieves a much smaller variance compared to the clustering method. We also show the model prediction, which agrees well at high densities where $\dw > 2$ and appears to come back into agreement at very low densities, $\dw \simeq 0.1$. 	
		
			\begin{figure*}
				\graphic{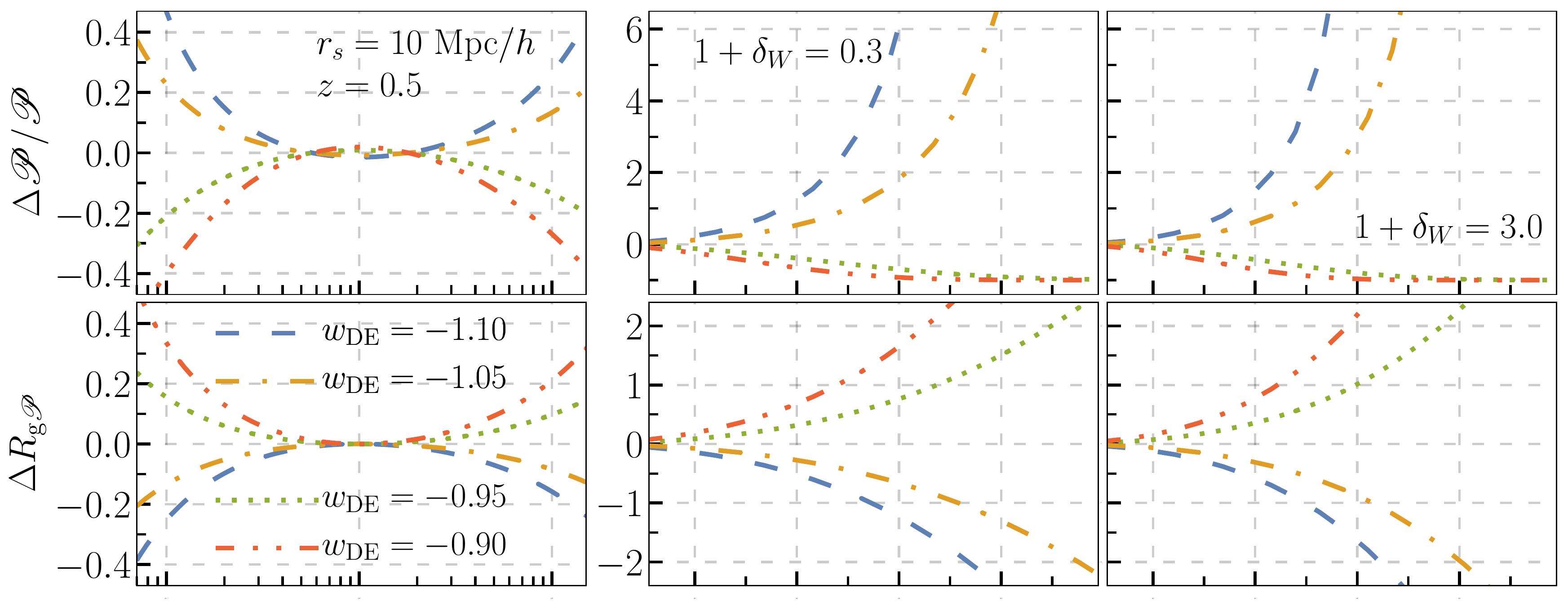}{1.} \\
				\vspace{-0.18in}		
				~ ~ ~ ~ \rule{0.942\textwidth}{0.9pt}
				\graphic{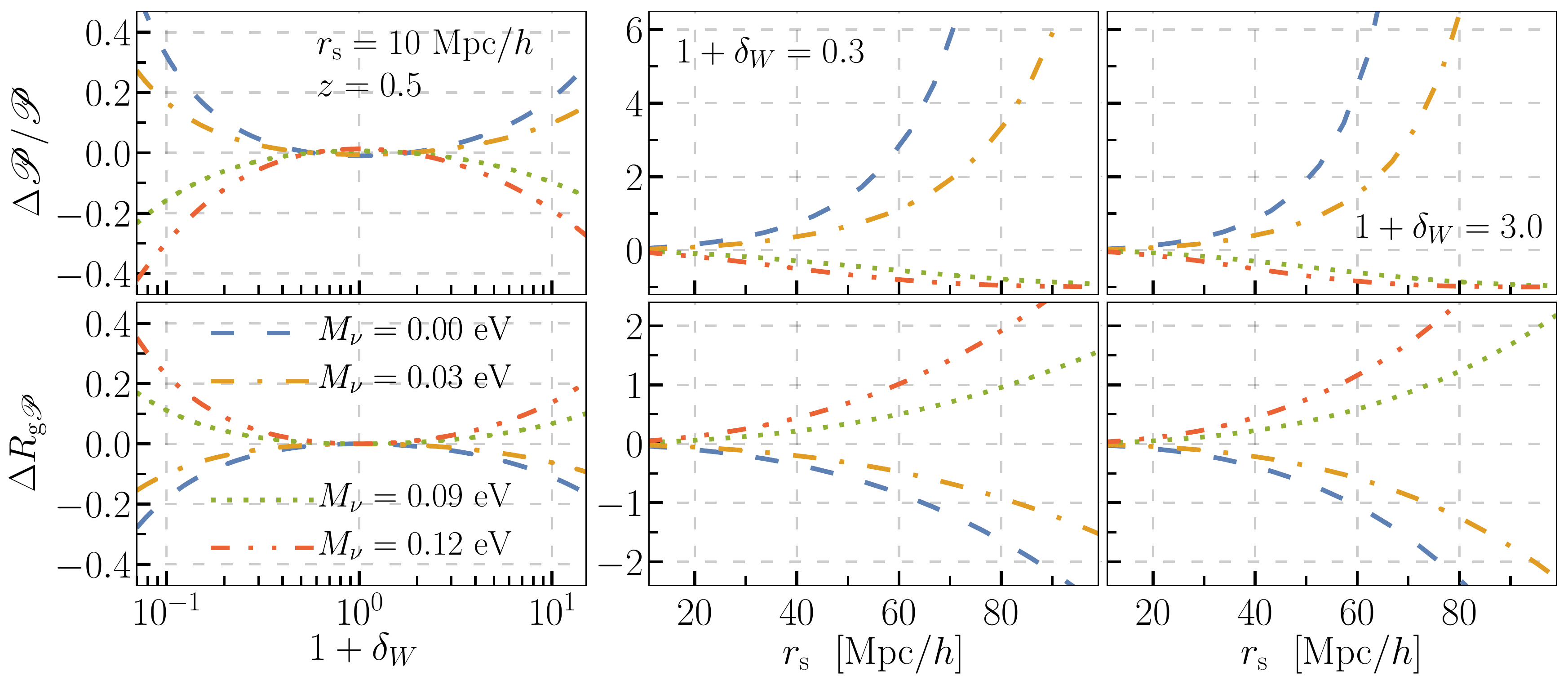}{1.}	
				\caption{Dependence of the PDF and its response on cosmological parameters at redshift $z=0.5$. The top two rows show our model calculations varying $w_\de$, the dark energy equation of state. The top row shows the fractional difference of the PDF compared with a fiducial cosmology where $w_\de=-1$. All other cosmological parameters are taken from the Planck 2018 best-fit values. The second row shows the PDF growth response compared with the fiducial cosmology. The left column shows the PDF and growth response as a function of density at fixed smoothing scale, while the middle and right columns show the dependence on the smoothing scale at fixed densities. The bottom two rows are the same set of plots, but varying the total neutrino mass $M_\nu$ while keeping the amount of cold dark matter and baryons fixed, as well as fixing the primordial power spectrum amplitude. The amount of dark energy is adjusted to satisfy the budget equation with vanishing curvature. The fiducial cosmology in this case has $M_{\nu} = 0.06\ \mathrm{eV}$, which corresponds to the minimal mass normal ordering and the central value for the allowed range with three degenerate neutrinos.}
				\label{fig:cosdep}
			\end{figure*}
			
		\section{Sensitivity to Cosmological Parameters}
			\label{sec:params}
		
			The shape of the PDF contains information about cosmological parameters. However, much of what goes into predicting the full shape (the EdS mapping from late-time to early-time densities, the aspherical prefactor) does not depend strongly on cosmology. The separate universe response of the PDF indicates how the shape of the PDF depends on $\Om$. We can interpret a long-wavelength mode as a local modulation in the value of $\Om$ throughout the universe, which locally affects the dynamics of gravitational clustering. 
				
			In Fig.~\ref{fig:cosdep}, we give examples of the sensitivity of the PDF and its growth response  to changes in cosmological parameters. The growth response is computed via the model in Eq.~(\ref{eq:Rgpdf}), which we have seen is in excellent agreement with simulations (Fig.~\ref{fig:epdfs}). The top two rows show effects of varying only the equation of state for dark energy, $w_\de = P_{\de}/\rho_{\de}$, where $P_{\de}$ is the dark energy pressure and $\rho_{\de}$ is the dark energy density. We have plotted the fractional difference in the PDF with respect to a fiducial cosmology in the top row, while the second row shows the difference in the growth response compared with the fiducial cosmology. Parameters for the fiducial cosmology were taken from the best-fit values of Planck 2018 \cite{Aghanim:2018eyx}. The left column shows the PDF and its growth response at a fixed smoothing scale, $\rs = 10~\mpc/h$. The middle and right columns show the effect of varying the smoothing scale at a fixed density. Similarly, the bottom two rows in Fig.~\ref{fig:cosdep} demonstrate the effects of varying the sum of neutrino masses, denoted $M_\nu$, for three degenerate neutrino species. In this case, the reference cosmology corresponds to the center of the range $0 < M_\nu < 0.12~\mathrm{eV}$ for three degenerate neutrinos. Note, the actual allowed range is $0.06~\mathrm{eV}<M_\nu<0.12~\mathrm{eV}$ \cite{Aghanim:2018eyx}, where the lower bound is based on observations of neutrino oscillations \cite{Zyla:2020zbs}, but we have included model calculations with lower masses for illustration. The upper bound increases if the neutrino mass hierarchy is assumed to have normal ($0.6~\mathrm{eV}<M_\nu<0.15~\mathrm{eV}$) or inverted ($0.1~\mathrm{eV}<M_\nu<0.17~\mathrm{eV}$) ordering \cite{RoyChoudhury:2019hls, Stocker:2020nsx}.
		
			Decreasing the equation of state, so that $w_\de < -1$, while keeping all other parameters fixed delays the onset of accelerated expansion. This enhances clustering, which leads to broader PDFs but diminishes the PDF response. Unlike the separate universe response, which is stronger at low density than at high density, the effect of $w_\de$ on the PDF is symmetric around $\log(1 +\dw) = 0$ (on a $\log$-scale).
		
			The effect of the neutrino mass is to introduce a free-streaming scale for large-scale clustering. Above this free-streaming scale, neutrino perturbations cluster and grow just like dark matter perturbations. Below this scale, the free streaming smooths out neutrino densities, so their perturbations decay away, which results in suppressed clustering for the dark matter. Increasing the neutrino mass leads to a suppression of clustering on small scales, which can be seen in Fig.~\ref{fig:cosdep}.  Here, we have fixed the amounts of dark matter and baryons ($\Oc$ and $\Ob$), and  the amplitude of the primordial power spectrum ($A_\mathrm{s}$), while adjusting the amount to dark energy ($\OL$) to maintain vanishing spatial curvature with the additional nonrelativistic matter from the massive neutrinos. This changes $\sigma_8$ and $\Om=\Oc+\Ob+\Onu$ for the different values of the neutrino mass.

			Since the effect of varying cosmological parameters on the shape of the PDF and its growth response are determined by the extent to which the parameter enhances or diminishes clustering, there appears to be a strong degeneracy between cosmological parameters if you only consider a single smoothing scale. However, the amount of increase in $M_\nu$ that is required to compensate for a decrease in $w_\de$, for example, will be different on different smoothing scales. By considering several different smoothing scales, it may be possible to break the parameter degeneracy and improve constraints using the shape of the PDF and its clustering, measured from observations.
		
		\section{Conclusion}
			\label{sec:con}
	
			In this paper, we have extended the separate universe formalism to the one-point statistics of the smoothed density field. The presence of long-wavelength density perturbations affects the clustering of small-scale density perturbations, which has a dynamical effect on the shape of the one-point statistics. This effect can be linearly characterized by the separate universe response of the density PDF. We measured this response and provided a model for it based on the power spectrum and its separate universe response. We also validated our simulation measurements, along with the separate universe consistency relation for the density PDF, by comparing to the linear clustering of the PDF measured in subboxes in simulations with the global cosmology.
		
			The model we presented is based on the collapse (expansion) of spherical overdensities (underdensities) in EdS. This model gives excellent agreement with our simulation results for the growth response of the PDF over the full range of densities and redshifts considered. The model gives poor reproduction of the shape of the PDF itself, which leads to disagreement with the nondynamical contributions to the PDF clustering. As shown in \cite{Ivanov:2018lcg}, this model can be improved by considering aspherical corrections, and since these corrections lead to accurate reproduction of the PDF, it would likely recover the nondynamical response terms. However, this calculation is quiet involved, and the results are insensitive to cosmology. Importantly, we have shown that these nondynamical terms can be accurately measured directly from data and subtracted from the PDF clustering, isolating the dynamical growth response, which is accurately predicted in the spherical model.
		
			The growth response of the PDF quantifies how sensitive regions of different small-scale densities are to shifts in $\Om$. In Sec.~\ref{sec:params}, we presented the effects of varying other cosmological parameters, such as the dark energy equation of state and the sum of neutrino masses, including how these effects differ when the PDF and its response are measured on different smoothing scales. We showed how considering the full shape of the PDF and its growth response at several smoothing scales may help break parameter degeneracies and improve constraints.
		
			The work presented here focused on a $\lcdm$ cosmology, for which the growth of long-wavelength perturbations is scale independent. As shown in Fig.~\ref{fig:local_clustering}, the linear clustering of the PDF is also scale independent in the linear regime, as is expected from the scale-independent growth. More general cosmologies that include primordial non-Gaussianity or massive neutrinos lead to scale-dependent effects on large scales, which would cause the clustering to be $k$ dependent in the linear regime, similar to the effect of scale-dependent bias.
			
			The growth response of the PDF indicates the sensitivity different regions of the density field have to shifts in cosmology, and may provide a useful way for obtaining optimal constraints on cosmological parameters. For example, marked correlation functions have been proposed as a way of boosting the information extracted from specific regions of the density field, such as voids \cite{White:2016yhs, Massara:2020pli}. Marked correlation functions are measured by first acting on observed densities with a nonlinear transformation and then computing the N-point statistics of the transformed data. Using the separate universe response of the PDF, we can motivate optimal choices for the form of the nonlinear transformation used to measure the marked correlation functions, which may yield improvement on parameter constraints. The studies presented in this paper are just the first steps towards developing the position-dependent PDF as a cosmological tool. Important follow-up work is to develop methods to extract the PDF response from data in the form of galaxy counts or weak lensing maps and compare these more complicated observables to our predictions. We leave this to future work.

	\acknowledgments 
	
	We would like to thank Mikhail Ivanov for assisting us with the model calculations. Results in this paper were obtained using the high-performance computing system at the Institute for Advanced Computational Science at Stony Brook University. Our figures were made using the P\textsc{ython} package M\textsc{atplotlib} \cite{Hunter:2007ouj}, and many of our calculation were carried out using S\textsc{ci}P\textsc{y} \cite{Virtanen:2019joe} and N\textsc{um}P\textsc{y} \cite{Harris:2020xlr}. D.J. is supported by Grants No. NSF PHY-1620628 and No. DOE DE-SC0017848. M.L. is supported by Grant No. DOE DE-SC0017848.
			
	\bibliography{PositionDependentMatterDensityPDF.bib}

\begin{thebibliography}{62}%
\makeatletter
\providecommand \@ifxundefined [1]{%
 \@ifx{#1\undefined}
}%
\providecommand \@ifnum [1]{%
 \ifnum #1\expandafter \@firstoftwo
 \else \expandafter \@secondoftwo
 \fi
}%
\providecommand \@ifx [1]{%
 \ifx #1\expandafter \@firstoftwo
 \else \expandafter \@secondoftwo
 \fi
}%
\providecommand \natexlab [1]{#1}%
\providecommand \enquote  [1]{``#1''}%
\providecommand \bibnamefont  [1]{#1}%
\providecommand \bibfnamefont [1]{#1}%
\providecommand \citenamefont [1]{#1}%
\providecommand \href@noop [0]{\@secondoftwo}%
\providecommand \href [0]{\begingroup \@sanitize@url \@href}%
\providecommand \@href[1]{\@@startlink{#1}\@@href}%
\providecommand \@@href[1]{\endgroup#1\@@endlink}%
\providecommand \@sanitize@url [0]{\catcode `\\12\catcode `\$12\catcode
  `\&12\catcode `\#12\catcode `\^12\catcode `\_12\catcode `\%12\relax}%
\providecommand \@@startlink[1]{}%
\providecommand \@@endlink[0]{}%
\providecommand \url  [0]{\begingroup\@sanitize@url \@url }%
\providecommand \@url [1]{\endgroup\@href {#1}{\urlprefix }}%
\providecommand \urlprefix  [0]{URL }%
\providecommand \Eprint [0]{\href }%
\providecommand \doibase [0]{http://dx.doi.org/}%
\providecommand \selectlanguage [0]{\@gobble}%
\providecommand \bibinfo  [0]{\@secondoftwo}%
\providecommand \bibfield  [0]{\@secondoftwo}%
\providecommand \translation [1]{[#1]}%
\providecommand \BibitemOpen [0]{}%
\providecommand \bibitemStop [0]{}%
\providecommand \bibitemNoStop [0]{.\EOS\space}%
\providecommand \EOS [0]{\spacefactor3000\relax}%
\providecommand \BibitemShut  [1]{\csname bibitem#1\endcsname}%
\let\auto@bib@innerbib\@empty
%</preamble>
\bibitem [{\citenamefont {Ivezi\'c}\ \emph {et~al.}(2019)\citenamefont
  {Ivezi\'c} \emph {et~al.}}]{Ivezic:2008fe}%
  \BibitemOpen
  \bibfield  {author} {\bibinfo {author} {\bibfnamefont {{\v Z}.}~\bibnamefont
  {Ivezi\'c}} \emph {et~al.} (\bibinfo {collaboration} {LSST}),\ }\href
  {\doibase 10.3847/1538-4357/ab042c} {\bibfield  {journal} {\bibinfo
  {journal} {Astrophys. J.}\ }\textbf {\bibinfo {volume} {873}},\ \bibinfo
  {pages} {111} (\bibinfo {year} {2019})},\ \Eprint
  {http://arxiv.org/abs/0805.2366} {arXiv:0805.2366 [astro-ph]} \BibitemShut
  {NoStop}%
\bibitem [{\citenamefont {Laureijs}\ \emph {et~al.}(2011)\citenamefont
  {Laureijs} \emph {et~al.}}]{Laureijs:2011gra}%
  \BibitemOpen
  \bibfield  {author} {\bibinfo {author} {\bibfnamefont {R.}~\bibnamefont
  {Laureijs}} \emph {et~al.} (\bibinfo {collaboration} {EUCLID}),\ }\href@noop
  {} {\  (\bibinfo {year} {2011})},\ \Eprint {http://arxiv.org/abs/1110.3193}
  {arXiv:1110.3193 [astro-ph.CO]} \BibitemShut {NoStop}%
\bibitem [{\citenamefont {Spergel}\ \emph {et~al.}(2015)\citenamefont {Spergel}
  \emph {et~al.}}]{Spergel:2015sza}%
  \BibitemOpen
  \bibfield  {author} {\bibinfo {author} {\bibfnamefont {D.}~\bibnamefont
  {Spergel}} \emph {et~al.},\ }\href@noop {} {\  (\bibinfo {year} {2015})},\
  \Eprint {http://arxiv.org/abs/1503.03757} {arXiv:1503.03757 [astro-ph.IM]}
  \BibitemShut {NoStop}%
\bibitem [{\citenamefont {Aghamousa}\ \emph {et~al.}(2016)\citenamefont
  {Aghamousa} \emph {et~al.}}]{Aghamousa:2016zmz}%
  \BibitemOpen
  \bibfield  {author} {\bibinfo {author} {\bibfnamefont {A.}~\bibnamefont
  {Aghamousa}} \emph {et~al.} (\bibinfo {collaboration} {DESI}),\ }\href@noop
  {} {\  (\bibinfo {year} {2016})},\ \Eprint {http://arxiv.org/abs/1611.00036}
  {arXiv:1611.00036 [astro-ph.IM]} \BibitemShut {NoStop}%
\bibitem [{\citenamefont {{Hubble}}(1934)}]{Hubble_1934}%
  \BibitemOpen
  \bibfield  {author} {\bibinfo {author} {\bibfnamefont {E.}~\bibnamefont
  {{Hubble}}},\ }\href {\doibase 10.1086/143517} {\bibfield  {journal}
  {\bibinfo  {journal} {\apj}\ }\textbf {\bibinfo {volume} {79}},\ \bibinfo
  {pages} {8} (\bibinfo {year} {1934})}\BibitemShut {NoStop}%
\bibitem [{\citenamefont {Wild}\ \emph {et~al.}(2005)\citenamefont {Wild} \emph
  {et~al.}}]{Wild:2004me}%
  \BibitemOpen
  \bibfield  {author} {\bibinfo {author} {\bibfnamefont {V.}~\bibnamefont
  {Wild}} \emph {et~al.} (\bibinfo {collaboration} {2dFGRS}),\ }\href {\doibase
  10.1111/j.1365-2966.2004.08447.x} {\bibfield  {journal} {\bibinfo  {journal}
  {Mon. Not. Roy. Astron. Soc.}\ }\textbf {\bibinfo {volume} {356}},\ \bibinfo
  {pages} {247} (\bibinfo {year} {2005})},\ \Eprint
  {http://arxiv.org/abs/astro-ph/0404275} {arXiv:astro-ph/0404275} \BibitemShut
  {NoStop}%
\bibitem [{\citenamefont {Hurtado-Gil}\ \emph {et~al.}(2017)\citenamefont
  {Hurtado-Gil}, \citenamefont {Mart\'\i{}nez}, \citenamefont {Arnalte-Mur},
  \citenamefont {Pons-Border\'\i{}a}, \citenamefont {Pareja-Flores},\ and\
  \citenamefont {Paredes}}]{Hurtado-Gil:2017dbm}%
  \BibitemOpen
  \bibfield  {author} {\bibinfo {author} {\bibfnamefont {L.}~\bibnamefont
  {Hurtado-Gil}}, \bibinfo {author} {\bibfnamefont {V.}~\bibnamefont
  {Mart\'\i{}nez}}, \bibinfo {author} {\bibfnamefont {P.}~\bibnamefont
  {Arnalte-Mur}}, \bibinfo {author} {\bibfnamefont {M.}~\bibnamefont
  {Pons-Border\'\i{}a}}, \bibinfo {author} {\bibfnamefont {C.}~\bibnamefont
  {Pareja-Flores}}, \ and\ \bibinfo {author} {\bibfnamefont {S.}~\bibnamefont
  {Paredes}},\ }\href {\doibase 10.1051/0004-6361/201629097} {\bibfield
  {journal} {\bibinfo  {journal} {Astron. Astrophys.}\ }\textbf {\bibinfo
  {volume} {601}},\ \bibinfo {pages} {A40} (\bibinfo {year} {2017})},\ \Eprint
  {http://arxiv.org/abs/1703.01087} {arXiv:1703.01087 [astro-ph.CO]}
  \BibitemShut {NoStop}%
\bibitem [{\citenamefont {Repp}\ and\ \citenamefont
  {Szapudi}(2020{\natexlab{a}})}]{Repp:2020kfd}%
  \BibitemOpen
  \bibfield  {author} {\bibinfo {author} {\bibfnamefont {A.}~\bibnamefont
  {Repp}}\ and\ \bibinfo {author} {\bibfnamefont {I.}~\bibnamefont {Szapudi}},\
  }\href {\doibase 10.1093/mnrasl/slaa139} {\bibfield  {journal} {\bibinfo
  {journal} {Mon. Not. Roy. Astron. Soc.}\ }\textbf {\bibinfo {volume} {498}},\
  \bibinfo {pages} {L125} (\bibinfo {year} {2020}{\natexlab{a}})},\ \Eprint
  {http://arxiv.org/abs/2006.01146} {arXiv:2006.01146 [astro-ph.CO]}
  \BibitemShut {NoStop}%
\bibitem [{\citenamefont {Clerkin}\ \emph {et~al.}(2017)\citenamefont {Clerkin}
  \emph {et~al.}}]{Clerkin:2016kyr}%
  \BibitemOpen
  \bibfield  {author} {\bibinfo {author} {\bibfnamefont {L.}~\bibnamefont
  {Clerkin}} \emph {et~al.} (\bibinfo {collaboration} {DES}),\ }\href {\doibase
  10.1093/mnras/stw2106} {\bibfield  {journal} {\bibinfo  {journal} {Mon. Not.
  Roy. Astron. Soc.}\ }\textbf {\bibinfo {volume} {466}},\ \bibinfo {pages}
  {1444} (\bibinfo {year} {2017})},\ \Eprint {http://arxiv.org/abs/1605.02036}
  {arXiv:1605.02036 [astro-ph.CO]} \BibitemShut {NoStop}%
\bibitem [{\citenamefont {Gruen}\ \emph {et~al.}(2018)\citenamefont {Gruen}
  \emph {et~al.}}]{Gruen:2017xjj}%
  \BibitemOpen
  \bibfield  {author} {\bibinfo {author} {\bibfnamefont {D.}~\bibnamefont
  {Gruen}} \emph {et~al.} (\bibinfo {collaboration} {DES}),\ }\href {\doibase
  10.1103/PhysRevD.98.023507} {\bibfield  {journal} {\bibinfo  {journal} {Phys.
  Rev. D}\ }\textbf {\bibinfo {volume} {98}},\ \bibinfo {pages} {023507}
  (\bibinfo {year} {2018})},\ \Eprint {http://arxiv.org/abs/1710.05045}
  {arXiv:1710.05045 [astro-ph.CO]} \BibitemShut {NoStop}%
\bibitem [{\citenamefont {Bouchet}\ and\ \citenamefont
  {Hernquist}(1992)}]{Bouchet:1992hw}%
  \BibitemOpen
  \bibfield  {author} {\bibinfo {author} {\bibfnamefont {F.}~\bibnamefont
  {Bouchet}}\ and\ \bibinfo {author} {\bibfnamefont {L.}~\bibnamefont
  {Hernquist}},\ }\href {\doibase 10.1086/171970} {\bibfield  {journal}
  {\bibinfo  {journal} {Astrophys. J.}\ }\textbf {\bibinfo {volume} {400}},\
  \bibinfo {pages} {25} (\bibinfo {year} {1992})}\BibitemShut {NoStop}%
\bibitem [{\citenamefont {Kofman}\ \emph {et~al.}(1994)\citenamefont {Kofman},
  \citenamefont {Bertschinger}, \citenamefont {Gelb}, \citenamefont {Nusser},\
  and\ \citenamefont {Dekel}}]{Kofman:1993mx}%
  \BibitemOpen
  \bibfield  {author} {\bibinfo {author} {\bibfnamefont {L.}~\bibnamefont
  {Kofman}}, \bibinfo {author} {\bibfnamefont {E.}~\bibnamefont
  {Bertschinger}}, \bibinfo {author} {\bibfnamefont {J.~M.}\ \bibnamefont
  {Gelb}}, \bibinfo {author} {\bibfnamefont {A.}~\bibnamefont {Nusser}}, \ and\
  \bibinfo {author} {\bibfnamefont {A.}~\bibnamefont {Dekel}},\ }\href
  {\doibase 10.1086/173541} {\bibfield  {journal} {\bibinfo  {journal}
  {Astrophys. J.}\ }\textbf {\bibinfo {volume} {420}},\ \bibinfo {pages} {44}
  (\bibinfo {year} {1994})},\ \Eprint {http://arxiv.org/abs/astro-ph/9311028}
  {arXiv:astro-ph/9311028} \BibitemShut {NoStop}%
\bibitem [{\citenamefont {Gaztanaga}\ \emph {et~al.}(2000)\citenamefont
  {Gaztanaga}, \citenamefont {Fosalba},\ and\ \citenamefont
  {Elizalde}}]{Gaztanaga:1999er}%
  \BibitemOpen
  \bibfield  {author} {\bibinfo {author} {\bibfnamefont {E.}~\bibnamefont
  {Gaztanaga}}, \bibinfo {author} {\bibfnamefont {P.}~\bibnamefont {Fosalba}},
  \ and\ \bibinfo {author} {\bibfnamefont {E.}~\bibnamefont {Elizalde}},\
  }\href {\doibase 10.1086/309249} {\bibfield  {journal} {\bibinfo  {journal}
  {Astrophys. J.}\ }\textbf {\bibinfo {volume} {539}},\ \bibinfo {pages} {522}
  (\bibinfo {year} {2000})},\ \Eprint {http://arxiv.org/abs/astro-ph/9906296}
  {arXiv:astro-ph/9906296} \BibitemShut {NoStop}%
\bibitem [{\citenamefont {Betancort-Rijo}\ and\ \citenamefont
  {Lopez-Corredoira}(2002)}]{BetancortRijo:2001ge}%
  \BibitemOpen
  \bibfield  {author} {\bibinfo {author} {\bibfnamefont {J.}~\bibnamefont
  {Betancort-Rijo}}\ and\ \bibinfo {author} {\bibfnamefont {M.}~\bibnamefont
  {Lopez-Corredoira}},\ }\href {\doibase 10.1086/338328} {\bibfield  {journal}
  {\bibinfo  {journal} {Astrophys. J.}\ }\textbf {\bibinfo {volume} {566}},\
  \bibinfo {pages} {623} (\bibinfo {year} {2002})},\ \Eprint
  {http://arxiv.org/abs/astro-ph/0110624} {arXiv:astro-ph/0110624} \BibitemShut
  {NoStop}%
\bibitem [{\citenamefont {Kayo}\ \emph {et~al.}(2001)\citenamefont {Kayo},
  \citenamefont {Taruya},\ and\ \citenamefont {Suto}}]{Kayo:2001gu}%
  \BibitemOpen
  \bibfield  {author} {\bibinfo {author} {\bibfnamefont {I.}~\bibnamefont
  {Kayo}}, \bibinfo {author} {\bibfnamefont {A.}~\bibnamefont {Taruya}}, \ and\
  \bibinfo {author} {\bibfnamefont {Y.}~\bibnamefont {Suto}},\ }\href {\doibase
  10.1086/323227} {\bibfield  {journal} {\bibinfo  {journal} {Astrophys. J.}\
  }\textbf {\bibinfo {volume} {561}},\ \bibinfo {pages} {22} (\bibinfo {year}
  {2001})},\ \Eprint {http://arxiv.org/abs/astro-ph/0105218}
  {arXiv:astro-ph/0105218} \BibitemShut {NoStop}%
\bibitem [{\citenamefont {Repp}\ and\ \citenamefont
  {Szapudi}(2018)}]{Repp:2017znp}%
  \BibitemOpen
  \bibfield  {author} {\bibinfo {author} {\bibfnamefont {A.}~\bibnamefont
  {Repp}}\ and\ \bibinfo {author} {\bibfnamefont {I.}~\bibnamefont {Szapudi}},\
  }\href {\doibase 10.1093/mnras/stx2615} {\bibfield  {journal} {\bibinfo
  {journal} {Mon. Not. Roy. Astron. Soc.}\ }\textbf {\bibinfo {volume} {473}},\
  \bibinfo {pages} {3598} (\bibinfo {year} {2018})},\ \Eprint
  {http://arxiv.org/abs/1705.08015} {arXiv:1705.08015 [astro-ph.CO]}
  \BibitemShut {NoStop}%
\bibitem [{\citenamefont {Shin}\ \emph {et~al.}(2017)\citenamefont {Shin},
  \citenamefont {Kim}, \citenamefont {Pichon}, \citenamefont {Jeong},\ and\
  \citenamefont {Park}}]{Shin:2017cwu}%
  \BibitemOpen
  \bibfield  {author} {\bibinfo {author} {\bibfnamefont {J.}~\bibnamefont
  {Shin}}, \bibinfo {author} {\bibfnamefont {J.}~\bibnamefont {Kim}}, \bibinfo
  {author} {\bibfnamefont {C.}~\bibnamefont {Pichon}}, \bibinfo {author}
  {\bibfnamefont {D.}~\bibnamefont {Jeong}}, \ and\ \bibinfo {author}
  {\bibfnamefont {C.}~\bibnamefont {Park}},\ }\href {\doibase
  10.3847/1538-4357/aa74b9} {\bibfield  {journal} {\bibinfo  {journal}
  {Astrophys. J.}\ }\textbf {\bibinfo {volume} {843}},\ \bibinfo {pages} {73}
  (\bibinfo {year} {2017})},\ \Eprint {http://arxiv.org/abs/1705.06863}
  {arXiv:1705.06863 [astro-ph.CO]} \BibitemShut {NoStop}%
\bibitem [{\citenamefont {Klypin}\ \emph {et~al.}(2018)\citenamefont {Klypin},
  \citenamefont {Prada}, \citenamefont {Betancort-Rijo},\ and\ \citenamefont
  {Albareti}}]{Klypin:2017jjg}%
  \BibitemOpen
  \bibfield  {author} {\bibinfo {author} {\bibfnamefont {A.}~\bibnamefont
  {Klypin}}, \bibinfo {author} {\bibfnamefont {F.}~\bibnamefont {Prada}},
  \bibinfo {author} {\bibfnamefont {J.}~\bibnamefont {Betancort-Rijo}}, \ and\
  \bibinfo {author} {\bibfnamefont {F.~D.}\ \bibnamefont {Albareti}},\ }\href
  {\doibase 10.1093/mnras/sty2613} {\bibfield  {journal} {\bibinfo  {journal}
  {Mon. Not. Roy. Astron. Soc.}\ }\textbf {\bibinfo {volume} {481}},\ \bibinfo
  {pages} {4588} (\bibinfo {year} {2018})},\ \Eprint
  {http://arxiv.org/abs/1706.01909} {arXiv:1706.01909 [astro-ph.CO]}
  \BibitemShut {NoStop}%
\bibitem [{\citenamefont {Bernardeau}(1994)}]{Bernardeau:1994zd}%
  \BibitemOpen
  \bibfield  {author} {\bibinfo {author} {\bibfnamefont {F.}~\bibnamefont
  {Bernardeau}},\ }\href@noop {} {\bibfield  {journal} {\bibinfo  {journal}
  {Astron. Astrophys.}\ }\textbf {\bibinfo {volume} {291}},\ \bibinfo {pages}
  {697} (\bibinfo {year} {1994})},\ \Eprint
  {http://arxiv.org/abs/astro-ph/9403020} {arXiv:astro-ph/9403020} \BibitemShut
  {NoStop}%
\bibitem [{\citenamefont {Bernardeau}\ and\ \citenamefont
  {Kofman}(1995)}]{Bernardeau:1994aq}%
  \BibitemOpen
  \bibfield  {author} {\bibinfo {author} {\bibfnamefont {F.}~\bibnamefont
  {Bernardeau}}\ and\ \bibinfo {author} {\bibfnamefont {L.}~\bibnamefont
  {Kofman}},\ }\href {\doibase 10.1086/175542} {\bibfield  {journal} {\bibinfo
  {journal} {Astrophys. J.}\ }\textbf {\bibinfo {volume} {443}},\ \bibinfo
  {pages} {479} (\bibinfo {year} {1995})},\ \Eprint
  {http://arxiv.org/abs/astro-ph/9403028} {arXiv:astro-ph/9403028} \BibitemShut
  {NoStop}%
\bibitem [{\citenamefont {Valageas}(2002)}]{Valageas:2001zr}%
  \BibitemOpen
  \bibfield  {author} {\bibinfo {author} {\bibfnamefont {P.}~\bibnamefont
  {Valageas}},\ }\href {\doibase 10.1051/0004-6361:20011663} {\bibfield
  {journal} {\bibinfo  {journal} {Astron. Astrophys.}\ }\textbf {\bibinfo
  {volume} {382}},\ \bibinfo {pages} {412} (\bibinfo {year} {2002})},\ \Eprint
  {http://arxiv.org/abs/astro-ph/0107126} {arXiv:astro-ph/0107126} \BibitemShut
  {NoStop}%
\bibitem [{\citenamefont {Ohta}\ \emph {et~al.}(2003)\citenamefont {Ohta},
  \citenamefont {Kayo},\ and\ \citenamefont {Taruya}}]{Ohta:2003zc}%
  \BibitemOpen
  \bibfield  {author} {\bibinfo {author} {\bibfnamefont {Y.}~\bibnamefont
  {Ohta}}, \bibinfo {author} {\bibfnamefont {I.}~\bibnamefont {Kayo}}, \ and\
  \bibinfo {author} {\bibfnamefont {A.}~\bibnamefont {Taruya}},\ }\href
  {\doibase 10.1086/374375} {\bibfield  {journal} {\bibinfo  {journal}
  {Astrophys. J.}\ }\textbf {\bibinfo {volume} {589}},\ \bibinfo {pages} {1}
  (\bibinfo {year} {2003})},\ \Eprint {http://arxiv.org/abs/astro-ph/0301567}
  {arXiv:astro-ph/0301567} \BibitemShut {NoStop}%
\bibitem [{\citenamefont {Lam}\ and\ \citenamefont {Sheth}(2008)}]{Lam:2007qw}%
  \BibitemOpen
  \bibfield  {author} {\bibinfo {author} {\bibfnamefont {T.}~\bibnamefont
  {Lam}}\ and\ \bibinfo {author} {\bibfnamefont {R.}~\bibnamefont {Sheth}},\
  }\href {\doibase 10.1111/j.1365-2966.2008.13038.x} {\bibfield  {journal}
  {\bibinfo  {journal} {Mon. Not. Roy. Astron. Soc.}\ }\textbf {\bibinfo
  {volume} {386}},\ \bibinfo {pages} {407} (\bibinfo {year} {2008})},\ \Eprint
  {http://arxiv.org/abs/0711.5029} {arXiv:0711.5029 [astro-ph]} \BibitemShut
  {NoStop}%
\bibitem [{\citenamefont {Uhlemann}\ \emph {et~al.}(2016)\citenamefont
  {Uhlemann}, \citenamefont {Codis}, \citenamefont {Pichon}, \citenamefont
  {Bernardeau},\ and\ \citenamefont {Reimberg}}]{Uhlemann:2015npz}%
  \BibitemOpen
  \bibfield  {author} {\bibinfo {author} {\bibfnamefont {C.}~\bibnamefont
  {Uhlemann}}, \bibinfo {author} {\bibfnamefont {S.}~\bibnamefont {Codis}},
  \bibinfo {author} {\bibfnamefont {C.}~\bibnamefont {Pichon}}, \bibinfo
  {author} {\bibfnamefont {F.}~\bibnamefont {Bernardeau}}, \ and\ \bibinfo
  {author} {\bibfnamefont {P.}~\bibnamefont {Reimberg}},\ }\href {\doibase
  10.1093/mnras/stw1074} {\bibfield  {journal} {\bibinfo  {journal} {Mon. Not.
  Roy. Astron. Soc.}\ }\textbf {\bibinfo {volume} {460}},\ \bibinfo {pages}
  {1529} (\bibinfo {year} {2016})},\ \Eprint {http://arxiv.org/abs/1512.05793}
  {arXiv:1512.05793 [astro-ph.CO]} \BibitemShut {NoStop}%
\bibitem [{\citenamefont {Ivanov}\ \emph {et~al.}(2019)\citenamefont {Ivanov},
  \citenamefont {Kaurov},\ and\ \citenamefont {Sibiryakov}}]{Ivanov:2018lcg}%
  \BibitemOpen
  \bibfield  {author} {\bibinfo {author} {\bibfnamefont {M.~M.}\ \bibnamefont
  {Ivanov}}, \bibinfo {author} {\bibfnamefont {A.~A.}\ \bibnamefont {Kaurov}},
  \ and\ \bibinfo {author} {\bibfnamefont {S.}~\bibnamefont {Sibiryakov}},\
  }\href {\doibase 10.1088/1475-7516/2019/03/009} {\bibfield  {journal}
  {\bibinfo  {journal} {JCAP}\ }\textbf {\bibinfo {volume} {03}},\ \bibinfo
  {pages} {009} (\bibinfo {year} {2019})},\ \Eprint
  {http://arxiv.org/abs/1811.07913} {arXiv:1811.07913 [astro-ph.CO]}
  \BibitemShut {NoStop}%
\bibitem [{\citenamefont {Friedrich}\ \emph {et~al.}(2020)\citenamefont
  {Friedrich}, \citenamefont {Uhlemann}, \citenamefont {Villaescusa-Navarro},
  \citenamefont {Baldauf}, \citenamefont {Manera},\ and\ \citenamefont
  {Nishimichi}}]{Friedrich:2019byw}%
  \BibitemOpen
  \bibfield  {author} {\bibinfo {author} {\bibfnamefont {O.}~\bibnamefont
  {Friedrich}}, \bibinfo {author} {\bibfnamefont {C.}~\bibnamefont {Uhlemann}},
  \bibinfo {author} {\bibfnamefont {F.}~\bibnamefont {Villaescusa-Navarro}},
  \bibinfo {author} {\bibfnamefont {T.}~\bibnamefont {Baldauf}}, \bibinfo
  {author} {\bibfnamefont {M.}~\bibnamefont {Manera}}, \ and\ \bibinfo {author}
  {\bibfnamefont {T.}~\bibnamefont {Nishimichi}},\ }\href {\doibase
  10.1093/mnras/staa2160} {\bibfield  {journal} {\bibinfo  {journal} {Mon. Not.
  Roy. Astron. Soc.}\ }\textbf {\bibinfo {volume} {498}},\ \bibinfo {pages}
  {464} (\bibinfo {year} {2020})},\ \Eprint {http://arxiv.org/abs/1912.06621}
  {arXiv:1912.06621 [astro-ph.CO]} \BibitemShut {NoStop}%
\bibitem [{\citenamefont {Uhlemann}\ \emph {et~al.}(2020)\citenamefont
  {Uhlemann}, \citenamefont {Friedrich}, \citenamefont {Villaescusa-Navarro},
  \citenamefont {Banerjee},\ and\ \citenamefont {Codis}}]{Uhlemann:2019gni}%
  \BibitemOpen
  \bibfield  {author} {\bibinfo {author} {\bibfnamefont {C.}~\bibnamefont
  {Uhlemann}}, \bibinfo {author} {\bibfnamefont {O.}~\bibnamefont {Friedrich}},
  \bibinfo {author} {\bibfnamefont {F.}~\bibnamefont {Villaescusa-Navarro}},
  \bibinfo {author} {\bibfnamefont {A.}~\bibnamefont {Banerjee}}, \ and\
  \bibinfo {author} {\bibfnamefont {S.}~\bibnamefont {Codis}},\ }\href
  {\doibase 10.1093/mnras/staa1155} {\bibfield  {journal} {\bibinfo  {journal}
  {Monthly Notices of the Royal Astronomical Society}\ }\textbf {\bibinfo
  {volume} {495}},\ \bibinfo {pages} {4006} (\bibinfo {year} {2020})},\ \Eprint
  {http://arxiv.org/abs/1911.11158} {arXiv:1911.11158 [astro-ph.CO]}
  \BibitemShut {NoStop}%
\bibitem [{\citenamefont {Repp}\ and\ \citenamefont
  {Szapudi}(2020{\natexlab{b}})}]{Repp:2020etr}%
  \BibitemOpen
  \bibfield  {author} {\bibinfo {author} {\bibfnamefont {A.}~\bibnamefont
  {Repp}}\ and\ \bibinfo {author} {\bibfnamefont {I.}~\bibnamefont {Szapudi}},\
  }\href {\doibase 10.1093/mnras/staa3237} {\bibfield  {journal} {\bibinfo
  {journal} {Mon. Not. Roy. Astron. Soc.}\ }\textbf {\bibinfo {volume} {500}},\
  \bibinfo {pages} {3631} (\bibinfo {year} {2020}{\natexlab{b}})},\ \Eprint
  {http://arxiv.org/abs/2007.00011} {arXiv:2007.00011 [astro-ph.CO]}
  \BibitemShut {NoStop}%
\bibitem [{\citenamefont {Maldacena}(2003)}]{Maldacena:2002vr}%
  \BibitemOpen
  \bibfield  {author} {\bibinfo {author} {\bibfnamefont {J.~M.}\ \bibnamefont
  {Maldacena}},\ }\href {\doibase 10.1088/1126-6708/2003/05/013} {\bibfield
  {journal} {\bibinfo  {journal} {JHEP}\ }\textbf {\bibinfo {volume} {05}},\
  \bibinfo {pages} {013} (\bibinfo {year} {2003})},\ \Eprint
  {http://arxiv.org/abs/astro-ph/0210603} {arXiv:astro-ph/0210603} \BibitemShut
  {NoStop}%
\bibitem [{\citenamefont {Creminelli}\ and\ \citenamefont
  {Zaldarriaga}(2004)}]{Creminelli:2004yq}%
  \BibitemOpen
  \bibfield  {author} {\bibinfo {author} {\bibfnamefont {P.}~\bibnamefont
  {Creminelli}}\ and\ \bibinfo {author} {\bibfnamefont {M.}~\bibnamefont
  {Zaldarriaga}},\ }\href {\doibase 10.1088/1475-7516/2004/10/006} {\bibfield
  {journal} {\bibinfo  {journal} {JCAP}\ }\textbf {\bibinfo {volume} {10}},\
  \bibinfo {pages} {006} (\bibinfo {year} {2004})},\ \Eprint
  {http://arxiv.org/abs/astro-ph/0407059} {arXiv:astro-ph/0407059} \BibitemShut
  {NoStop}%
\bibitem [{\citenamefont {Li}\ \emph {et~al.}(2014{\natexlab{a}})\citenamefont
  {Li}, \citenamefont {Hu},\ and\ \citenamefont {Takada}}]{Li:2014sga}%
  \BibitemOpen
  \bibfield  {author} {\bibinfo {author} {\bibfnamefont {Y.}~\bibnamefont
  {Li}}, \bibinfo {author} {\bibfnamefont {W.}~\bibnamefont {Hu}}, \ and\
  \bibinfo {author} {\bibfnamefont {M.}~\bibnamefont {Takada}},\ }\href
  {\doibase 10.1103/PhysRevD.89.083519} {\bibfield  {journal} {\bibinfo
  {journal} {Phys. Rev. D}\ }\textbf {\bibinfo {volume} {89}},\ \bibinfo
  {pages} {083519} (\bibinfo {year} {2014}{\natexlab{a}})},\ \Eprint
  {http://arxiv.org/abs/1401.0385} {arXiv:1401.0385 [astro-ph.CO]} \BibitemShut
  {NoStop}%
\bibitem [{\citenamefont {Chiang}\ \emph {et~al.}(2014)\citenamefont {Chiang},
  \citenamefont {Wagner}, \citenamefont {Schmidt},\ and\ \citenamefont
  {Komatsu}}]{Chiang:2014oga}%
  \BibitemOpen
  \bibfield  {author} {\bibinfo {author} {\bibfnamefont {C.-T.}\ \bibnamefont
  {Chiang}}, \bibinfo {author} {\bibfnamefont {C.}~\bibnamefont {Wagner}},
  \bibinfo {author} {\bibfnamefont {F.}~\bibnamefont {Schmidt}}, \ and\
  \bibinfo {author} {\bibfnamefont {E.}~\bibnamefont {Komatsu}},\ }\href
  {\doibase 10.1088/1475-7516/2014/05/048} {\bibfield  {journal} {\bibinfo
  {journal} {JCAP}\ }\textbf {\bibinfo {volume} {05}},\ \bibinfo {pages} {048}
  (\bibinfo {year} {2014})},\ \Eprint {http://arxiv.org/abs/1403.3411}
  {arXiv:1403.3411 [astro-ph.CO]} \BibitemShut {NoStop}%
\bibitem [{\citenamefont {Li}\ \emph {et~al.}(2016)\citenamefont {Li},
  \citenamefont {Hu},\ and\ \citenamefont {Takada}}]{Li:2015jsz}%
  \BibitemOpen
  \bibfield  {author} {\bibinfo {author} {\bibfnamefont {Y.}~\bibnamefont
  {Li}}, \bibinfo {author} {\bibfnamefont {W.}~\bibnamefont {Hu}}, \ and\
  \bibinfo {author} {\bibfnamefont {M.}~\bibnamefont {Takada}},\ }\href
  {\doibase 10.1103/PhysRevD.93.063507} {\bibfield  {journal} {\bibinfo
  {journal} {Phys. Rev. D}\ }\textbf {\bibinfo {volume} {93}},\ \bibinfo
  {pages} {063507} (\bibinfo {year} {2016})},\ \Eprint
  {http://arxiv.org/abs/1511.01454} {arXiv:1511.01454 [astro-ph.CO]}
  \BibitemShut {NoStop}%
\bibitem [{\citenamefont {Baldauf}\ \emph {et~al.}(2016)\citenamefont
  {Baldauf}, \citenamefont {Seljak}, \citenamefont {Senatore},\ and\
  \citenamefont {Zaldarriaga}}]{Baldauf:2015vio}%
  \BibitemOpen
  \bibfield  {author} {\bibinfo {author} {\bibfnamefont {T.}~\bibnamefont
  {Baldauf}}, \bibinfo {author} {\bibfnamefont {U.}~\bibnamefont {Seljak}},
  \bibinfo {author} {\bibfnamefont {L.}~\bibnamefont {Senatore}}, \ and\
  \bibinfo {author} {\bibfnamefont {M.}~\bibnamefont {Zaldarriaga}},\ }\href
  {\doibase 10.1088/1475-7516/2016/09/007} {\bibfield  {journal} {\bibinfo
  {journal} {JCAP}\ }\textbf {\bibinfo {volume} {09}},\ \bibinfo {pages} {007}
  (\bibinfo {year} {2016})},\ \Eprint {http://arxiv.org/abs/1511.01465}
  {arXiv:1511.01465 [astro-ph.CO]} \BibitemShut {NoStop}%
\bibitem [{\citenamefont {Chiang}\ \emph {et~al.}(2018)\citenamefont {Chiang},
  \citenamefont {Hu}, \citenamefont {Li},\ and\ \citenamefont
  {Loverde}}]{Chiang:2017vuk}%
  \BibitemOpen
  \bibfield  {author} {\bibinfo {author} {\bibfnamefont {C.-T.}\ \bibnamefont
  {Chiang}}, \bibinfo {author} {\bibfnamefont {W.}~\bibnamefont {Hu}}, \bibinfo
  {author} {\bibfnamefont {Y.}~\bibnamefont {Li}}, \ and\ \bibinfo {author}
  {\bibfnamefont {M.}~\bibnamefont {Loverde}},\ }\href {\doibase
  10.1103/PhysRevD.97.123526} {\bibfield  {journal} {\bibinfo  {journal} {Phys.
  Rev. D}\ }\textbf {\bibinfo {volume} {97}},\ \bibinfo {pages} {123526}
  (\bibinfo {year} {2018})},\ \Eprint {http://arxiv.org/abs/1710.01310}
  {arXiv:1710.01310 [astro-ph.CO]} \BibitemShut {NoStop}%
\bibitem [{\citenamefont {Chiang}\ \emph {et~al.}(2016)\citenamefont {Chiang},
  \citenamefont {Li}, \citenamefont {Hu},\ and\ \citenamefont
  {Loverde}}]{Chiang:2016vxa}%
  \BibitemOpen
  \bibfield  {author} {\bibinfo {author} {\bibfnamefont {C.-T.}\ \bibnamefont
  {Chiang}}, \bibinfo {author} {\bibfnamefont {Y.}~\bibnamefont {Li}}, \bibinfo
  {author} {\bibfnamefont {W.}~\bibnamefont {Hu}}, \ and\ \bibinfo {author}
  {\bibfnamefont {M.}~\bibnamefont {Loverde}},\ }\href {\doibase
  10.1103/PhysRevD.94.123502} {\bibfield  {journal} {\bibinfo  {journal} {Phys.
  Rev. D}\ }\textbf {\bibinfo {volume} {94}},\ \bibinfo {pages} {123502}
  (\bibinfo {year} {2016})},\ \Eprint {http://arxiv.org/abs/1609.01701}
  {arXiv:1609.01701 [astro-ph.CO]} \BibitemShut {NoStop}%
\bibitem [{\citenamefont {Jamieson}\ and\ \citenamefont
  {Loverde}(2019{\natexlab{a}})}]{Jamieson:2018biz}%
  \BibitemOpen
  \bibfield  {author} {\bibinfo {author} {\bibfnamefont {D.}~\bibnamefont
  {Jamieson}}\ and\ \bibinfo {author} {\bibfnamefont {M.}~\bibnamefont
  {Loverde}},\ }\href {\doibase 10.1103/PhysRevD.100.023516} {\bibfield
  {journal} {\bibinfo  {journal} {Phys. Rev. D}\ }\textbf {\bibinfo {volume}
  {100}},\ \bibinfo {pages} {023516} (\bibinfo {year} {2019}{\natexlab{a}})},\
  \Eprint {http://arxiv.org/abs/1812.08765} {arXiv:1812.08765 [astro-ph.CO]}
  \BibitemShut {NoStop}%
\bibitem [{\citenamefont {Chan}\ \emph {et~al.}(2020)\citenamefont {Chan},
  \citenamefont {Li}, \citenamefont {Biagetti},\ and\ \citenamefont
  {Hamaus}}]{Chan:2019yzq}%
  \BibitemOpen
  \bibfield  {author} {\bibinfo {author} {\bibfnamefont {K.~C.}\ \bibnamefont
  {Chan}}, \bibinfo {author} {\bibfnamefont {Y.}~\bibnamefont {Li}}, \bibinfo
  {author} {\bibfnamefont {M.}~\bibnamefont {Biagetti}}, \ and\ \bibinfo
  {author} {\bibfnamefont {N.}~\bibnamefont {Hamaus}},\ }\href {\doibase
  10.3847/1538-4357/ab64ec} {\bibfield  {journal} {\bibinfo  {journal}
  {Astrophys. J.}\ }\textbf {\bibinfo {volume} {889}},\ \bibinfo {pages} {89}
  (\bibinfo {year} {2020})},\ \Eprint {http://arxiv.org/abs/1909.03736}
  {arXiv:1909.03736 [astro-ph.CO]} \BibitemShut {NoStop}%
\bibitem [{\citenamefont {Jamieson}\ and\ \citenamefont
  {Loverde}(2019{\natexlab{b}})}]{Jamieson:2019dmp}%
  \BibitemOpen
  \bibfield  {author} {\bibinfo {author} {\bibfnamefont {D.}~\bibnamefont
  {Jamieson}}\ and\ \bibinfo {author} {\bibfnamefont {M.}~\bibnamefont
  {Loverde}},\ }\href {\doibase 10.1103/PhysRevD.100.123528} {\bibfield
  {journal} {\bibinfo  {journal} {Phys. Rev. D}\ }\textbf {\bibinfo {volume}
  {100}},\ \bibinfo {pages} {123528} (\bibinfo {year} {2019}{\natexlab{b}})},\
  \Eprint {http://arxiv.org/abs/1909.05313} {arXiv:1909.05313 [astro-ph.CO]}
  \BibitemShut {NoStop}%
\bibitem [{\citenamefont {Meerburg}\ \emph {et~al.}(2019)\citenamefont
  {Meerburg} \emph {et~al.}}]{Meerburg:2019qqi}%
  \BibitemOpen
  \bibfield  {author} {\bibinfo {author} {\bibfnamefont {P.~D.}\ \bibnamefont
  {Meerburg}} \emph {et~al.},\ }\href@noop {} {\  (\bibinfo {year} {2019})},\
  \Eprint {http://arxiv.org/abs/1903.04409} {arXiv:1903.04409 [astro-ph.CO]}
  \BibitemShut {NoStop}%
\bibitem [{\citenamefont {Dvorkin}\ \emph {et~al.}(2019)\citenamefont {Dvorkin}
  \emph {et~al.}}]{Dvorkin:2019jgs}%
  \BibitemOpen
  \bibfield  {author} {\bibinfo {author} {\bibfnamefont {C.}~\bibnamefont
  {Dvorkin}} \emph {et~al.},\ }\href@noop {} {\  (\bibinfo {year} {2019})},\
  \Eprint {http://arxiv.org/abs/1903.03689} {arXiv:1903.03689 [astro-ph.CO]}
  \BibitemShut {NoStop}%
\bibitem [{\citenamefont {Slosar}\ \emph {et~al.}(2019)\citenamefont {Slosar}
  \emph {et~al.}}]{Slosar:2019flp}%
  \BibitemOpen
  \bibfield  {author} {\bibinfo {author} {\bibfnamefont {A.}~\bibnamefont
  {Slosar}} \emph {et~al.},\ }\href@noop {} {\  (\bibinfo {year} {2019})},\
  \Eprint {http://arxiv.org/abs/1903.12016} {arXiv:1903.12016 [astro-ph.CO]}
  \BibitemShut {NoStop}%
\bibitem [{\citenamefont {White}(2016)}]{White:2016yhs}%
  \BibitemOpen
  \bibfield  {author} {\bibinfo {author} {\bibfnamefont {M.}~\bibnamefont
  {White}},\ }\href {\doibase 10.1088/1475-7516/2016/11/057} {\bibfield
  {journal} {\bibinfo  {journal} {JCAP}\ }\textbf {\bibinfo {volume} {11}},\
  \bibinfo {pages} {057} (\bibinfo {year} {2016})},\ \Eprint
  {http://arxiv.org/abs/1609.08632} {arXiv:1609.08632 [astro-ph.CO]}
  \BibitemShut {NoStop}%
\bibitem [{\citenamefont {Massara}\ \emph {et~al.}(2021)\citenamefont
  {Massara}, \citenamefont {Villaescusa-Navarro}, \citenamefont {Ho},
  \citenamefont {Dalal},\ and\ \citenamefont {Spergel}}]{Massara:2020pli}%
  \BibitemOpen
  \bibfield  {author} {\bibinfo {author} {\bibfnamefont {E.}~\bibnamefont
  {Massara}}, \bibinfo {author} {\bibfnamefont {F.}~\bibnamefont
  {Villaescusa-Navarro}}, \bibinfo {author} {\bibfnamefont {S.}~\bibnamefont
  {Ho}}, \bibinfo {author} {\bibfnamefont {N.}~\bibnamefont {Dalal}}, \ and\
  \bibinfo {author} {\bibfnamefont {D.~N.}\ \bibnamefont {Spergel}},\ }\href
  {\doibase 10.1103/PhysRevLett.126.011301} {\bibfield  {journal} {\bibinfo
  {journal} {Phys. Rev. Lett.}\ }\textbf {\bibinfo {volume} {126}},\ \bibinfo
  {pages} {011301} (\bibinfo {year} {2021})},\ \Eprint
  {http://arxiv.org/abs/2001.11024} {arXiv:2001.11024 [astro-ph.CO]}
  \BibitemShut {NoStop}%
\bibitem [{\citenamefont {Blas}\ \emph {et~al.}(2011)\citenamefont {Blas},
  \citenamefont {Lesgourgues},\ and\ \citenamefont {Tram}}]{Blas:2011rf}%
  \BibitemOpen
  \bibfield  {author} {\bibinfo {author} {\bibfnamefont {D.}~\bibnamefont
  {Blas}}, \bibinfo {author} {\bibfnamefont {J.}~\bibnamefont {Lesgourgues}}, \
  and\ \bibinfo {author} {\bibfnamefont {T.}~\bibnamefont {Tram}},\ }\href
  {\doibase 10.1088/1475-7516/2011/07/034} {\bibfield  {journal} {\bibinfo
  {journal} {JCAP}\ }\textbf {\bibinfo {volume} {07}},\ \bibinfo {pages} {034}
  (\bibinfo {year} {2011})},\ \Eprint {http://arxiv.org/abs/1104.2933}
  {arXiv:1104.2933 [astro-ph.CO]} \BibitemShut {NoStop}%
\bibitem [{\citenamefont {Bernardeau}\ \emph {et~al.}(2002)\citenamefont
  {Bernardeau}, \citenamefont {Colombi}, \citenamefont {Gaztanaga},\ and\
  \citenamefont {Scoccimarro}}]{Bernardeau:2001qr}%
  \BibitemOpen
  \bibfield  {author} {\bibinfo {author} {\bibfnamefont {F.}~\bibnamefont
  {Bernardeau}}, \bibinfo {author} {\bibfnamefont {S.}~\bibnamefont {Colombi}},
  \bibinfo {author} {\bibfnamefont {E.}~\bibnamefont {Gaztanaga}}, \ and\
  \bibinfo {author} {\bibfnamefont {R.}~\bibnamefont {Scoccimarro}},\ }\href
  {\doibase 10.1016/S0370-1573(02)00135-7} {\bibfield  {journal} {\bibinfo
  {journal} {Phys. Rept.}\ }\textbf {\bibinfo {volume} {367}},\ \bibinfo
  {pages} {1} (\bibinfo {year} {2002})},\ \Eprint
  {http://arxiv.org/abs/astro-ph/0112551} {arXiv:astro-ph/0112551} \BibitemShut
  {NoStop}%
\bibitem [{\citenamefont {Sirko}(2005)}]{Sirko:2005uz}%
  \BibitemOpen
  \bibfield  {author} {\bibinfo {author} {\bibfnamefont {E.}~\bibnamefont
  {Sirko}},\ }\href {\doibase 10.1086/497090} {\bibfield  {journal} {\bibinfo
  {journal} {Astrophys. J.}\ }\textbf {\bibinfo {volume} {634}},\ \bibinfo
  {pages} {728} (\bibinfo {year} {2005})},\ \Eprint
  {http://arxiv.org/abs/astro-ph/0503106} {arXiv:astro-ph/0503106} \BibitemShut
  {NoStop}%
\bibitem [{\citenamefont {Li}\ \emph {et~al.}(2014{\natexlab{b}})\citenamefont
  {Li}, \citenamefont {Hu},\ and\ \citenamefont {Takada}}]{Li:2014jra}%
  \BibitemOpen
  \bibfield  {author} {\bibinfo {author} {\bibfnamefont {Y.}~\bibnamefont
  {Li}}, \bibinfo {author} {\bibfnamefont {W.}~\bibnamefont {Hu}}, \ and\
  \bibinfo {author} {\bibfnamefont {M.}~\bibnamefont {Takada}},\ }\href
  {\doibase 10.1103/PhysRevD.90.103530} {\bibfield  {journal} {\bibinfo
  {journal} {Phys. Rev. D}\ }\textbf {\bibinfo {volume} {90}},\ \bibinfo
  {pages} {103530} (\bibinfo {year} {2014}{\natexlab{b}})},\ \Eprint
  {http://arxiv.org/abs/1408.1081} {arXiv:1408.1081 [astro-ph.CO]} \BibitemShut
  {NoStop}%
\bibitem [{\citenamefont {Takada}\ and\ \citenamefont
  {Hu}(2013)}]{Takada:2013wfa}%
  \BibitemOpen
  \bibfield  {author} {\bibinfo {author} {\bibfnamefont {M.}~\bibnamefont
  {Takada}}\ and\ \bibinfo {author} {\bibfnamefont {W.}~\bibnamefont {Hu}},\
  }\href {\doibase 10.1103/PhysRevD.87.123504} {\bibfield  {journal} {\bibinfo
  {journal} {Phys. Rev. D}\ }\textbf {\bibinfo {volume} {87}},\ \bibinfo
  {pages} {123504} (\bibinfo {year} {2013})},\ \Eprint
  {http://arxiv.org/abs/1302.6994} {arXiv:1302.6994 [astro-ph.CO]} \BibitemShut
  {NoStop}%
\bibitem [{\citenamefont {Valageas}(2014)}]{Valageas:2013zda}%
  \BibitemOpen
  \bibfield  {author} {\bibinfo {author} {\bibfnamefont {P.}~\bibnamefont
  {Valageas}},\ }\href {\doibase 10.1103/PhysRevD.89.123522} {\bibfield
  {journal} {\bibinfo  {journal} {Phys. Rev. D}\ }\textbf {\bibinfo {volume}
  {89}},\ \bibinfo {pages} {123522} (\bibinfo {year} {2014})},\ \Eprint
  {http://arxiv.org/abs/1311.4286} {arXiv:1311.4286 [astro-ph.CO]} \BibitemShut
  {NoStop}%
\bibitem [{\citenamefont {Springel}(2005)}]{Springel:2005mi}%
  \BibitemOpen
  \bibfield  {author} {\bibinfo {author} {\bibfnamefont {V.}~\bibnamefont
  {Springel}},\ }\href {\doibase 10.1111/j.1365-2966.2005.09655.x} {\bibfield
  {journal} {\bibinfo  {journal} {Mon. Not. Roy. Astron. Soc.}\ }\textbf
  {\bibinfo {volume} {364}},\ \bibinfo {pages} {1105} (\bibinfo {year}
  {2005})},\ \Eprint {http://arxiv.org/abs/astro-ph/0505010}
  {arXiv:astro-ph/0505010 [astro-ph]} \BibitemShut {NoStop}%
%%CITATION = ASTRO-PH/0505010;%%
\bibitem [{\citenamefont {Wagner}\ \emph {et~al.}(2015)\citenamefont {Wagner},
  \citenamefont {Schmidt}, \citenamefont {Chiang},\ and\ \citenamefont
  {Komatsu}}]{Wagner:2014aka}%
  \BibitemOpen
  \bibfield  {author} {\bibinfo {author} {\bibfnamefont {C.}~\bibnamefont
  {Wagner}}, \bibinfo {author} {\bibfnamefont {F.}~\bibnamefont {Schmidt}},
  \bibinfo {author} {\bibfnamefont {C.-T.}\ \bibnamefont {Chiang}}, \ and\
  \bibinfo {author} {\bibfnamefont {E.}~\bibnamefont {Komatsu}},\ }\href
  {\doibase 10.1093/mnrasl/slu187} {\bibfield  {journal} {\bibinfo  {journal}
  {Mon. Not. Roy. Astron. Soc.}\ }\textbf {\bibinfo {volume} {448}},\ \bibinfo
  {pages} {L11} (\bibinfo {year} {2015})},\ \Eprint
  {http://arxiv.org/abs/1409.6294} {arXiv:1409.6294 [astro-ph.CO]} \BibitemShut
  {NoStop}%
\bibitem [{\citenamefont {Coles}\ and\ \citenamefont
  {Jones}(1991)}]{Coles:1991if}%
  \BibitemOpen
  \bibfield  {author} {\bibinfo {author} {\bibfnamefont {P.}~\bibnamefont
  {Coles}}\ and\ \bibinfo {author} {\bibfnamefont {B.}~\bibnamefont {Jones}},\
  }\href {\doibase 10.1093/mnras/248.1.1} {\bibfield  {journal} {\bibinfo
  {journal} {mnras}\ }\textbf {\bibinfo {volume} {248}},\ \bibinfo {pages} {1}
  (\bibinfo {year} {1991})}\BibitemShut {NoStop}%
\bibitem [{\citenamefont {Silverman}(1986)}]{Silverman_1986}%
  \BibitemOpen
  \bibfield  {author} {\bibinfo {author} {\bibfnamefont {B.~W.}\ \bibnamefont
  {Silverman}},\ }\href@noop {} {\emph {\bibinfo {title} {Density Estimation
  for Statistics and Data Analysis}}}\ (\bibinfo  {publisher} {Chapman \&
  Hall},\ \bibinfo {address} {London},\ \bibinfo {year} {1986})\BibitemShut
  {NoStop}%
\bibitem [{\citenamefont {{Frigo}}\ and\ \citenamefont
  {{Johnson}}(2005)}]{FFTW05}%
  \BibitemOpen
  \bibfield  {author} {\bibinfo {author} {\bibfnamefont {M.}~\bibnamefont
  {{Frigo}}}\ and\ \bibinfo {author} {\bibfnamefont {S.~G.}\ \bibnamefont
  {{Johnson}}},\ }\href {\doibase 10.1109/JPROC.2004.840301} {\bibfield
  {journal} {\bibinfo  {journal} {Proceedings of the IEEE}\ }\textbf {\bibinfo
  {volume} {93}},\ \bibinfo {pages} {216} (\bibinfo {year} {2005})}\BibitemShut
  {NoStop}%
\bibitem [{\citenamefont {Aghanim}\ \emph {et~al.}(2020)\citenamefont {Aghanim}
  \emph {et~al.}}]{Aghanim:2018eyx}%
  \BibitemOpen
  \bibfield  {author} {\bibinfo {author} {\bibfnamefont {N.}~\bibnamefont
  {Aghanim}} \emph {et~al.} (\bibinfo {collaboration} {Planck}),\ }\href
  {\doibase 10.1051/0004-6361/201833910} {\bibfield  {journal} {\bibinfo
  {journal} {Astron. Astrophys.}\ }\textbf {\bibinfo {volume} {641}},\ \bibinfo
  {pages} {A6} (\bibinfo {year} {2020})},\ \Eprint
  {http://arxiv.org/abs/1807.06209} {arXiv:1807.06209 [astro-ph.CO]}
  \BibitemShut {NoStop}%
\bibitem [{\citenamefont {Zyla}\ \emph {et~al.}(2020)\citenamefont {Zyla} \emph
  {et~al.}}]{Zyla:2020zbs}%
  \BibitemOpen
  \bibfield  {author} {\bibinfo {author} {\bibfnamefont {P.}~\bibnamefont
  {Zyla}} \emph {et~al.} (\bibinfo {collaboration} {Particle Data Group}),\
  }\href {\doibase 10.1093/ptep/ptaa104} {\bibfield  {journal} {\bibinfo
  {journal} {PTEP}\ }\textbf {\bibinfo {volume} {2020}},\ \bibinfo {pages}
  {083C01} (\bibinfo {year} {2020})}\BibitemShut {NoStop}%
\bibitem [{\citenamefont {Roy~Choudhury}\ and\ \citenamefont
  {Hannestad}(2020)}]{RoyChoudhury:2019hls}%
  \BibitemOpen
  \bibfield  {author} {\bibinfo {author} {\bibfnamefont {S.}~\bibnamefont
  {Roy~Choudhury}}\ and\ \bibinfo {author} {\bibfnamefont {S.}~\bibnamefont
  {Hannestad}},\ }\href {\doibase 10.1088/1475-7516/2020/07/037} {\bibfield
  {journal} {\bibinfo  {journal} {JCAP}\ }\textbf {\bibinfo {volume} {07}},\
  \bibinfo {pages} {037} (\bibinfo {year} {2020})},\ \Eprint
  {http://arxiv.org/abs/1907.12598} {arXiv:1907.12598 [astro-ph.CO]}
  \BibitemShut {NoStop}%
\bibitem [{\citenamefont {St\"ocker}\ \emph {et~al.}(2020)\citenamefont
  {St\"ocker} \emph {et~al.}}]{Stocker:2020nsx}%
  \BibitemOpen
  \bibfield  {author} {\bibinfo {author} {\bibfnamefont {P.}~\bibnamefont
  {St\"ocker}} \emph {et~al.} (\bibinfo {collaboration} {GAMBIT Cosmology
  Workgroup}),\ }\href@noop {} {\  (\bibinfo {year} {2020})},\ \Eprint
  {http://arxiv.org/abs/2009.03287} {arXiv:2009.03287 [astro-ph.CO]}
  \BibitemShut {NoStop}%
\bibitem [{\citenamefont {Hunter}(2007)}]{Hunter:2007ouj}%
  \BibitemOpen
  \bibfield  {author} {\bibinfo {author} {\bibfnamefont {J.~D.}\ \bibnamefont
  {Hunter}},\ }\href {\doibase 10.1109/MCSE.2007.55} {\bibfield  {journal}
  {\bibinfo  {journal} {Comput. Sci. Eng.}\ }\textbf {\bibinfo {volume} {9}},\
  \bibinfo {pages} {90} (\bibinfo {year} {2007})}\BibitemShut {NoStop}%
\bibitem [{\citenamefont {Virtanen}\ \emph {et~al.}(2020)\citenamefont
  {Virtanen} \emph {et~al.}}]{Virtanen:2019joe}%
  \BibitemOpen
  \bibfield  {author} {\bibinfo {author} {\bibfnamefont {P.}~\bibnamefont
  {Virtanen}} \emph {et~al.},\ }\href {\doibase 10.1038/s41592-019-0686-2}
  {\bibfield  {journal} {\bibinfo  {journal} {Nature Meth.}\ }\textbf {\bibinfo
  {volume} {17}},\ \bibinfo {pages} {261} (\bibinfo {year} {2020})},\ \Eprint
  {http://arxiv.org/abs/1907.10121} {arXiv:1907.10121 [cs.MS]} \BibitemShut
  {NoStop}%
\bibitem [{\citenamefont {Harris}\ \emph {et~al.}(2020)\citenamefont {Harris}
  \emph {et~al.}}]{Harris:2020xlr}%
  \BibitemOpen
  \bibfield  {author} {\bibinfo {author} {\bibfnamefont {C.~R.}\ \bibnamefont
  {Harris}} \emph {et~al.},\ }\href {\doibase 10.1038/s41586-020-2649-2}
  {\bibfield  {journal} {\bibinfo  {journal} {Nature}\ }\textbf {\bibinfo
  {volume} {585}},\ \bibinfo {pages} {357} (\bibinfo {year} {2020})},\ \Eprint
  {http://arxiv.org/abs/2006.10256} {arXiv:2006.10256 [cs.MS]} \BibitemShut
  {NoStop}%
\end{thebibliography}%
	
\end{document}